\documentclass[
	superscriptaddress,
	showpacs,
	preprintnumbers,
	amsmath,
	amssymb,
	twocolumn,
	aps,
	prapplied,
	]{revtex4-2}

\usepackage{graphicx}
\usepackage[utf8]{inputenc}
\usepackage[T1]{fontenc}
\usepackage{mathptmx}
\usepackage{siunitx}
\usepackage{xfrac}
\usepackage{xcolor}
\usepackage{bm}
\usepackage{amsmath}

\begin{document}

\title{Transverse magnetic routing of light emission in hybrid plasmonic-semiconductor nanostructures: Towards operation at room temperature} 

\author{Lars Klompmaker}
\email[]{lars.klompmaker@tu-dortmund.de}
\affiliation{Experimentelle Physik 2, Technische Universit\"at Dortmund, 44221 Dortmund, Germany}

\author{Alexander N. Poddubny}
\affiliation{Ioffe Institute, Russian Academy of Sciences, 194021 St. Petersburg, Russia}

\author{Ey\"up Yalcin}
\affiliation{Experimentelle Physik 2, Technische Universit\"at Dortmund, 44221 Dortmund, Germany}

\author{Leonid V. Litvin}
\affiliation{Raith GmbH, 44263 Dortmund, Germany}

\author{Ralf Jede}
\affiliation{Raith GmbH, 44263 Dortmund, Germany}

\author{Grzegorz Karczewski}
\affiliation{Institute of Physics, Polish Academy of Sciences, PL-02668 Warsaw, Poland}

\author{Sergij Chusnutdinow}
\affiliation{Institute of Physics, Polish Academy of Sciences, PL-02668 Warsaw, Poland}

\author{Tomasz Wojtowicz}
\affiliation{International Research Centre MagTop, Institute of Physics, Polish Academy of Sciences, PL-02668 Warsaw, Poland}

\author{Dmitri R. Yakovlev}
\affiliation{Experimentelle Physik 2, Technische Universit\"at Dortmund, 44221 Dortmund, Germany}
\affiliation{Ioffe Institute, Russian Academy of Sciences, 194021 St. Petersburg, Russia}

\author{Manfred Bayer}
\affiliation{Experimentelle Physik 2, Technische Universit\"at Dortmund, 44221 Dortmund, Germany}
\affiliation{Ioffe Institute, Russian Academy of Sciences, 194021 St. Petersburg, Russia}

\author{Ilya A. Akimov}
\affiliation{Experimentelle Physik 2, Technische Universit\"at Dortmund, 44221 Dortmund, Germany}
\affiliation{Ioffe Institute, Russian Academy of Sciences, 194021 St. Petersburg, Russia}

\date{\today}
\begin{abstract}
We study experimentally and theoretically the temperature dependence of transverse magnetic routing of light emission from hybrid plasmonic-semiconductor quantum well structures where the exciton emission from the quantum well is routed into surface plasmon polaritons propagating along a nearby semiconductor-metal interface. 
In II-VI and III-V direct band semiconductors the magnitude of routing is governed by the circular polarization of exciton optical transitions, that is induced by a magnetic field. 
For structures comprising a (Cd,Mn)Te/(Cd,Mg)Te diluted magnetic semiconductor quantum well we observe a strong directionality of the emission up to $15\%$ at low temperature of $20$~K and magnetic field of $485$~mT due to giant Zeeman splitting of holes mediated via the strong exchange interaction with Mn$^{2+}$ ions. 
For increasing temperatures towards room-temperature the magnetic susceptibility decreases and the directionality strongly decreases to $4\%$ at $T=45$~K.
We also propose an alternative design based on a non-magnetic (In,Ga)As/(In,Al)As quantum well structure, suitable for higher temperatures. 
According to our calculations, such structure can demonstrate emission directionality up to $5\%$ for temperatures below $200$~K and moderate magnetic fields of $1$~T.
\end{abstract}


\maketitle 

\section{Introduction}
Recent achievements in nanotechnology boosted rapid development of magnetophotonics -- an emerging field where a magnetic field is used to alter the optical response of nanophotonic structures, e.g.\ amplitude, phase and polarization of the electromagnetic wave transmitted through the structure \cite{Magnetophotonics-book}. 
It is now well established that spatial localization of light at the nanoscale leads to significant enhancement of magneto-optical effects which opens new opportunities for applications in optical communication and metrology~\cite{Kavokin-Vladimirova-1997,Granovsky-2005,Awschalom-2006, Pisarev-Rasing-2008, Vladimirova-2013}. 
Significant progress has been achieved by combination of noble metals with magnetic materials resulting in hybrid plasmonic structures, where the magnetic field induces a substantial modulation of optical spectra and their polarization in the vicinity of plasmonic resonances due to enhancement of the Faraday or Kerr magneto-optical effects \cite{Armelles-2013, Maccaferri-2020}. 
Here, particular interest is devoted to non-reciprocal intensity effects such as the transverse magneto-optical Kerr effect~\cite{Safarov-2001, Temnov-2010, Belotelov-2011, Kreilkamp-2013, Poddubny-BCS-2020} which is otherwise very small in homogeneous ferromagnetic films~\cite{Zvezdin-book, Krinchik-59, Martin-65}.

\begin{figure}
	\includegraphics[width=0.45\textwidth]{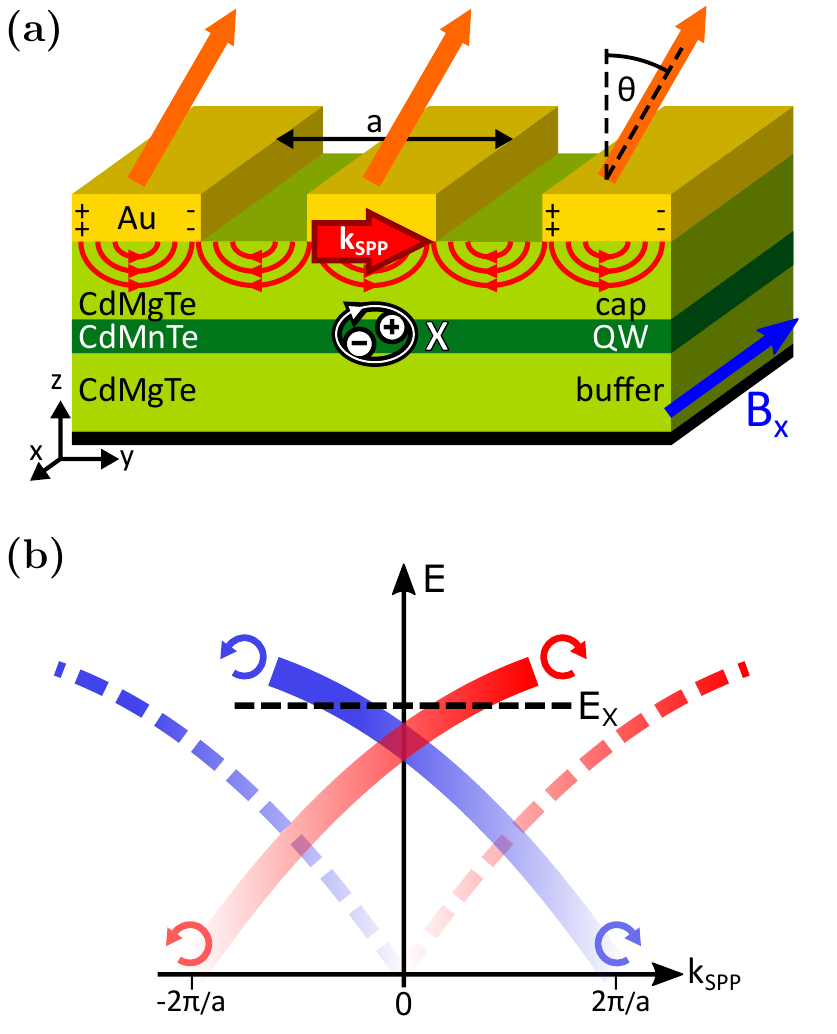}
	\caption{(a) Schematic presentation of the hybrid plasmonic-semiconductor quantum well structure used in the experiments and the processes involved in achieving directional emission. The in-plane external magnetic field $B_x$ induces elliptically polarized optical transitions of QW excitons in the $yz$-plane (exciton X is considered as a point emitter) due to the Zeeman effect. The excitons couple predominantly to surface plasmon polaritons (SPPs, in red) with the same helicity of polarization at the semiconductor/gold interface leading to directional SPP propagation. The metal grating is required for detection of SPPs in the far field at angle $\theta$ defined by the grating period $a$.
	(b) Dispersion diagram of SPPs with opposite wave vectors $\mathbf{k}_\mathrm{SPP}$ and elliptical polarization (shown in red/blue and circular arrows, respectively), due to spin-momentum locking. 
	Dashed and solid curves show the SPP dispersion for the homogeneous film and for the grating with period $a$, respectively.
	The exciton (X) couples predominantly to the SPP modes with the same polarization helicity.
	}
	\label{fig:1}
\end{figure}

An important feature of nanophotonic structures is the possibility to tailor the polarization properties of the photonics modes which can be used for new functionalities. 
For example, evanescent electromagnetic waves at an interface such as surface plasmon polaritons (SPPs) have peculiar polarization properties. 
Namely, the electric field is elliptically polarized in the plane containing the wave vector $\bm k_{\rm SPP}$ and the normal to the surface $\bm z$ ($yz$-plane in Fig.~\ref{fig:1}(a)), and the sign of ellipticity depends on the propagating direction, as shown by the circular arrows in Fig.~\ref{fig:1}(b). 
This effect is termed spin-momentum locking\cite{Spin-Momentum-Locking,Leuchs-2015}: Evanescent plasmons carry so-called transverse spin $\propto \bm k_{\rm SPP}\times \bm z$. 
Such optical spin fluxes and even more advanced polarization features due to coupling between the spin and orbital degrees of freedom of light can be used to establish directional coupling to circularly polarized dipoles in various structures~\cite{Kapitanova-2014, SO-2015, Chiral-Optics-Review, Dyakov-2020}. 
It has been demonstrated that the radiative recombination of excitons in quantum dots placed at certain positions in photonic crystals or guiding structures leads to emission into the desired direction locked to their spin polarization~\cite{Oulton-2013, Sollner-2015, Coles-2016}. 
The spin polarization of excitons can be induced by applying an external magnetic field~\cite{Dyakonov-book, Kossut-book}. 
Using this approach we have recently demonstrated transverse magnetic routing of light emission (TMRLE) for diluted magnetic semiconductor (Cd,Mn)Te/(Cd,Mg)Te quantum well (QW) structures where the directional emission of excitons in external magnetic field was substantially enhanced due to coupling with surface plasmon polaritons~\cite{FelixNature-2018}. 
It should be noted that low-dimensional semiconductor structures attract particular interest because non-reciprocal magneto-optical effects are resonantly enhanced in the vicinity of the exciton resonances~\cite{Godde-2013, Kotova-2019, Borovkova-2019}.

The enhancement of TMRLE in hybrid plasmonic-semiconductor QW structures is based on directional coupling between exciton and evanescent SPP waves in the vicinity of the metal-semiconductor interface. 
Since the semiconductor quantum well is located only a few tens of nm apart from the interface the electron-hole pairs (excitons) efficiently excite surface plasmons when recombining radiatively, as shown in Fig.~\ref{fig:1}(a). 
When the magnetic field $\mathbf{B}||\mathbf{x}$ is applied in the plane of the structure, the exciton optical transitions gain circular polarization degree $P_\mathrm{c}$ in the $yz$-plane, perpendicular to the magnetic field direction. 
The sign and magnitude of $P_\mathrm{c}$ depend on the direction and strength of the magnetic field $B_x$, respectively.  
Because of the spin-momentum locking effect for surface plasmons, elliptically polarized excitons are directionally coupled to either left- or right-propagating plasmons, depending on the transition ellipticity defined by the sign of $P_\mathrm{c}$. 
Phenomenologically, the emitted surface plasmons propagate predominantly along one of the directions perpendicular to the magnetic field given by the wavevector $\mathbf{k_{SPP}} \propto \mathbf{B}\times \mathbf{e_z}$ (see Fig.~\ref{fig:1}(a)).
SPP waves in the homogeneous semiconductor-metal interface are evanescent.
In order to couple them out into far field radiation the homogeneous metal film is substituted by a one-dimensional plasmonic grating with a period $a$.
The angle of emission into the far field $\theta$ is then determined by matching the in-plane component of the emitted light wave vector $\omega/c\sin\theta$ with the diffracted SPP vector, so that $\omega/c\sin\theta = \pm (k_\mathrm{SPP}(\omega)\pm 2\pi/a)$, where $\omega$ is the light frequency and $c$ the speed of light in vacuum. 
The structure is designed in such way that the crossing point of the diffracted SPP dispersion branches (when $k_\mathrm{SPP}(\omega)=2\pi/a$, thick curves in Fig.~\ref{fig:1}(b)) is close to the exciton resonance energy $E_\mathrm{X}$ (dashed line X in Fig.~\ref{fig:1}(b)). 
This type of structure is also very versatile because the photon energy of emission is determined by the exciton energy $E_\mathrm{X}$, which can be tuned by the parameters of the semiconductor heterostructure. 
In experiment, 60\% of emission directionality has been recently achieved in diluted magnetic semiconductor (DMS) structures based on (Cd,Mn)Te QWs covered with gold gratings~\cite{FelixNature-2018} in moderate magnetic fields of $2$ Tesla. However, this number corresponds to low temperatures of $\SI{2}{K}$.

In this study we focus on the temperature dependence of the emission directionality in various hybrid plasmonic-semiconductor QW structures. 
We demonstrate experimentally that at higher temperatures the effect gets weaker due to the decreasing magnetic permittivity of the DMS QW, thus leading to a lower degree of circular polarization $P_\mathrm{c}$ in agreement with our theoretical modeling.
Therefore, we propose an alternative type of device where the DMS is replaced by a non-magnetic (In,Ga,Al)As quantum well structure (i.e. without magnetic ions) with intrinsically large hole g-factor.
Our calculations for such structure result in a considerable directionality above 5\% for $B=1$~T that is independent from temperature in a large range up to $200$~K in the telecom wavelength range around $\SI{1600}{nm}$.

The paper is organized as follows: 
In Section~\ref{sec:Polarization} we examine the polarization of optical transitions in DMS QWs which is essential for the description of TMRLE in hybrid plasmonic-semiconductor QW structures. 
Section~\ref{sec:Experimental} describes the experimental details. 
The temperature dependence of TMRLE in (Cd,Mn)Te based structures is considered in Section~\ref{sec:TMRLE-DMS}. 
Finally, Section ~\ref{sec:highT} considers alternative (i.e. temperature independent), non-magnetic structures (without magnetic ions) for the realization of TMRLE across a wide temperature range.

\section{Polarization of optical interband transitions in QW structures subject to external magnetic field}
\label{sec:Polarization}

\begin{figure*}
	\includegraphics[width=0.85\textwidth]{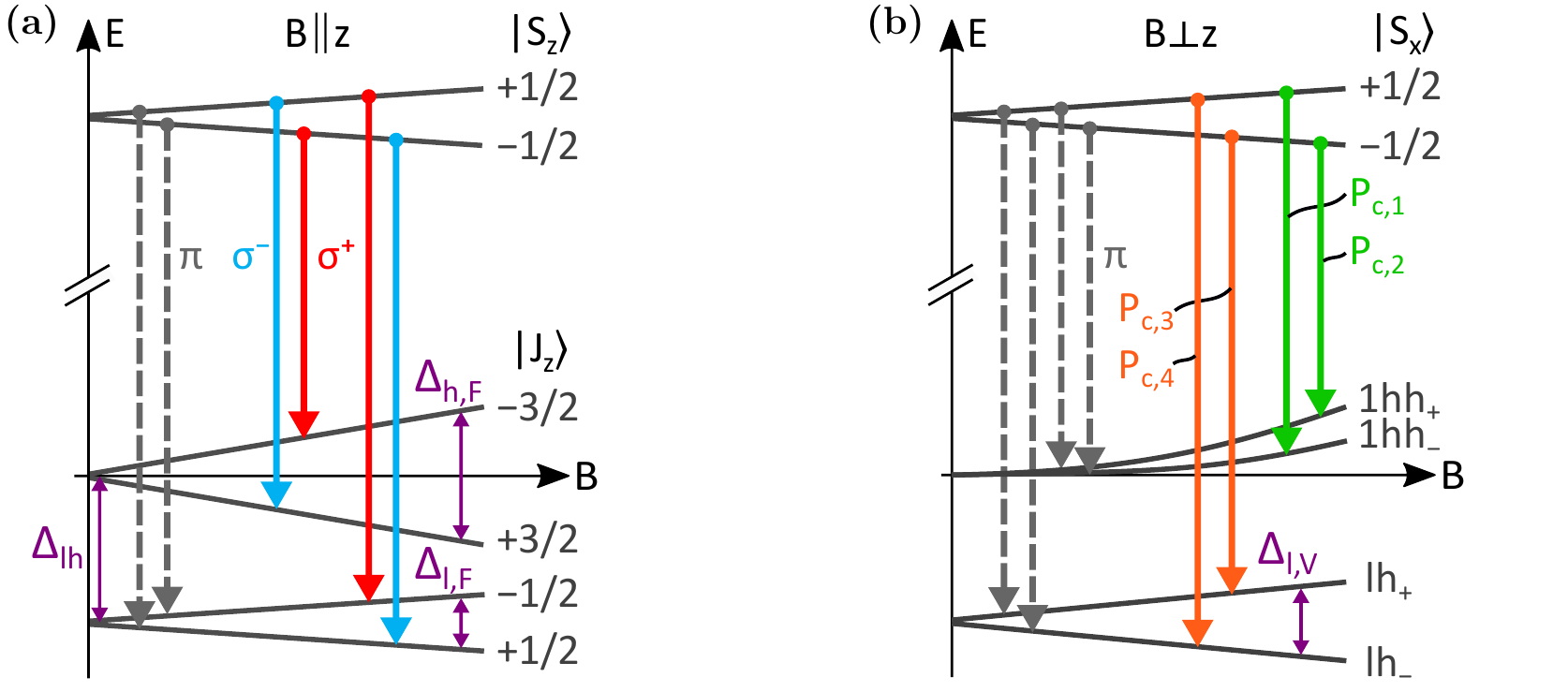}
	\caption{Zeeman splitting of conduction ($1e$) and valence (heavy-hole ($1hh$) and light-hole ($1lh$)) band states in longitudinal ($\mathbf{B}||z$) (a) and transverse ($\mathbf{B}\perp z$) (b) magnetic fields. The arrows show optical transitions, $E$ is the energy. 
	(a) The labels $\sigma^{\pm}$ and $\pi$ indicate the polarization of the transitions. 
	$\Delta_\mathrm{h,F}$ and $\Delta_\mathrm{l,F}$ are the Zeeman splittings of heavy- and light-holes in Faraday geometry, respectively, and $\Delta_\mathrm{lh}$ is the energy splitting between heavy- and light-holes at $B=0$.
	(b) $P_{\mathrm{c},i}$ ($i=1,2,3,4$) are the degrees of circular polarization in the $yz$-plane for the transitions according to Eq.~\eqref{eq:Pc0} and $\Delta_\mathrm{l,V}$ is the Zeeman splitting of light holes in Voigt geometry used to approximate $P_\mathrm{c}$ in Eq.~\eqref{eq:Pc}.	}
	\label{fig:pola1}
\end{figure*}

In this section we discuss the general selection rules for optical transitions in a quantum well, subjected to an external magnetic field. The main part of the theoretical analysis is presented in Refs.~\cite{FelixNature-2018,Borovkova-2019} and is based on Refs.~\cite{Zeeman-Voigt,Furdyna-review}. Here, we summarize the main results and underlying approximations. Furthermore, we extend the analysis of magnetic-field-induced circular polarization for higher energy interband optical transitions with light holes.

We describe the valence band states by the Luttinger Hamiltonian in spherical approximation and consider only linear terms in magnetic field in the Hamiltonian. For simplicity the effects of bulk inversion asymmetry such as magneto-spatial dispersion are ignored~\cite{Kotova-2019}. Furthermore,  fine structure splittings of exciton complexes due to the electron-hole exchange interaction are not taken into account~\cite{Ivchenko-book}.  Figure~\ref{fig:pola1} presents the scheme of conduction and valence band states split by the magnetic field. We consider only the lowest levels of quantization of electrons, heavy holes and light holes. Crucially, the light- and heavy-hole subbands are split even for zero magnetic field by the splitting $\Delta_\mathrm{lh}$, due to different quantization energies in the quantum well and strain due to lattice mismatch between the QW and barrier materials. Additionally, the application of the magnetic field leads to the splitting of the conduction and valence band states. This splitting depends strongly on the orientation of the magnetic field with respect to the QW, either along the growth direction [$\bm B\parallel z$, Fig.~\ref{fig:pola1}(a)] or perpendicular to it [$\bm B\perp z$, Fig.~\ref{fig:pola1}(b)]. 

In the Faraday geometry ($\bm B\parallel z$) both electron and hole states can be characterized by the projection of total angular momentum on the $z$ axis, $S_z$ for electrons and $J_z$ for holes. 
The Zeeman splitting for electron, light- and heavy-hole states is linear in magnetic field. 
For heavy- and light-holes, as indicated in Fig.~\ref{fig:pola1}(a), it is given by
\begin{equation}
	\Delta_{k,\rm F} = 2J_z g_{\mathrm h} \mu_{\mathrm B} B + \frac{2}{3} J_z xN_{0}\beta\left< S_{z}^\mathrm{Mn}\right>, 
\label{eq:Zeeman}
\end{equation}
where the index $k=\rm l$ for light holes with $J_z=\pm1/2$ and $k= \rm h$ for heavy holes with $J_z=\pm3/2$. The first term on the right hand side in Eq.~\eqref{eq:Zeeman} is responsible for the Zeeman splitting of valence band states with Land\'e $g$-factor of holes $g_{\rm h}$, while the second term is due to the exchange interaction between the holes and magnetic ${\rm Mn^{2+}}$ ions, which takes place in DMS quantum wells. Here, $\mu_\mathrm{B}$ is the Bohr magneton, $x$ is the Mn concentration, $N_{0}\beta = \SI{-0.88}{eV}$ is the exchange constant for the valence bands in CdMnTe and $\left< S_z^\mathrm{Mn}(B)\right>$ is the thermal average of the ${\rm Mn^{2+}}$ spin projection along ${\bf B}$~\cite{Ossau-1993}. This average can be described by the modified Brillouin function $B_S$ for $S = 5/2$:
\begin{equation}
	\left< S_{z}^\mathrm{Mn}(B)\right> = S_\mathrm{eff} B_{5/2}\left( \frac{5}{2} \frac{\mu_\mathrm{B} g_\mathrm{Mn} B}{k_\mathrm{B} (T_\mathrm{Mn}+T_0) } \right)\text{,}
	\label{eq:Brillouin}
\end{equation}
where $k_\mathrm{B}$ is the Boltzmann constant, $B$ is the externally applied magnetic field and $g_\mathrm{Mn} = 2$ is the $g$-factor of $\mathrm{Mn}^{2+}$ ions. $T_0$ is the effective temperature, which is a function of $x$, and $T_\mathrm{Mn}$ is the Mn-spin temperature.
The effective spin $S_{\mathrm{eff}}$ and effective temperature $T_0$ take into account that for $x > \SI{1}{\percent}$ the paramagnetic behavior is reduced due to the emerging antiferromagnetic exchange interaction between Mn-spins~\cite{Ossau-1993,Furdyna-review,Kossut-book}. For weak magnetic fields the Zeeman splittings are linear in $B$.

For (Cd,Mn)Te/(Cd,Mg)Te DMS quantum wells at low temperatures, the first term in Eq.~\eqref{eq:Zeeman} is negligible, as compared to the exchange interaction with the Mn$^{2+}$ ions. Thus, in this case it is enough to account only for the giant Zeeman splitting effect which is given by the second term~\cite{Zeeman-Voigt}. Nevertheless, in our model the Zeeman splitting is proportional to $J_z$ independent of its origin and therefore the identity $\Delta_\mathrm{h,F}=3\Delta_\mathrm{l,F}$ holds also in non-magnetic QW structures (see first term in Eq.~\eqref{eq:Zeeman}). In this case, however, the magnitude of the Zeeman splitting is given by the g-factor $g_{\rm h}$ which is independent of temperature.

The optical transitions with electric field in the QW plane are always fully circularly polarized ($\sigma^\pm$ arrows in Fig.~\ref{fig:pola1}(a)), irrespective of the applied magnetic field strength $B$. The linearly-polarized $\pi$-transitions in Fig.~\ref{fig:pola1}(a) correspond to the electric field of the electromagnetic wave $\bm E$ normal to the QW plane, $\bm E\parallel \bm B\parallel z$.

\begin{figure}
	\includegraphics[width=0.45\textwidth]{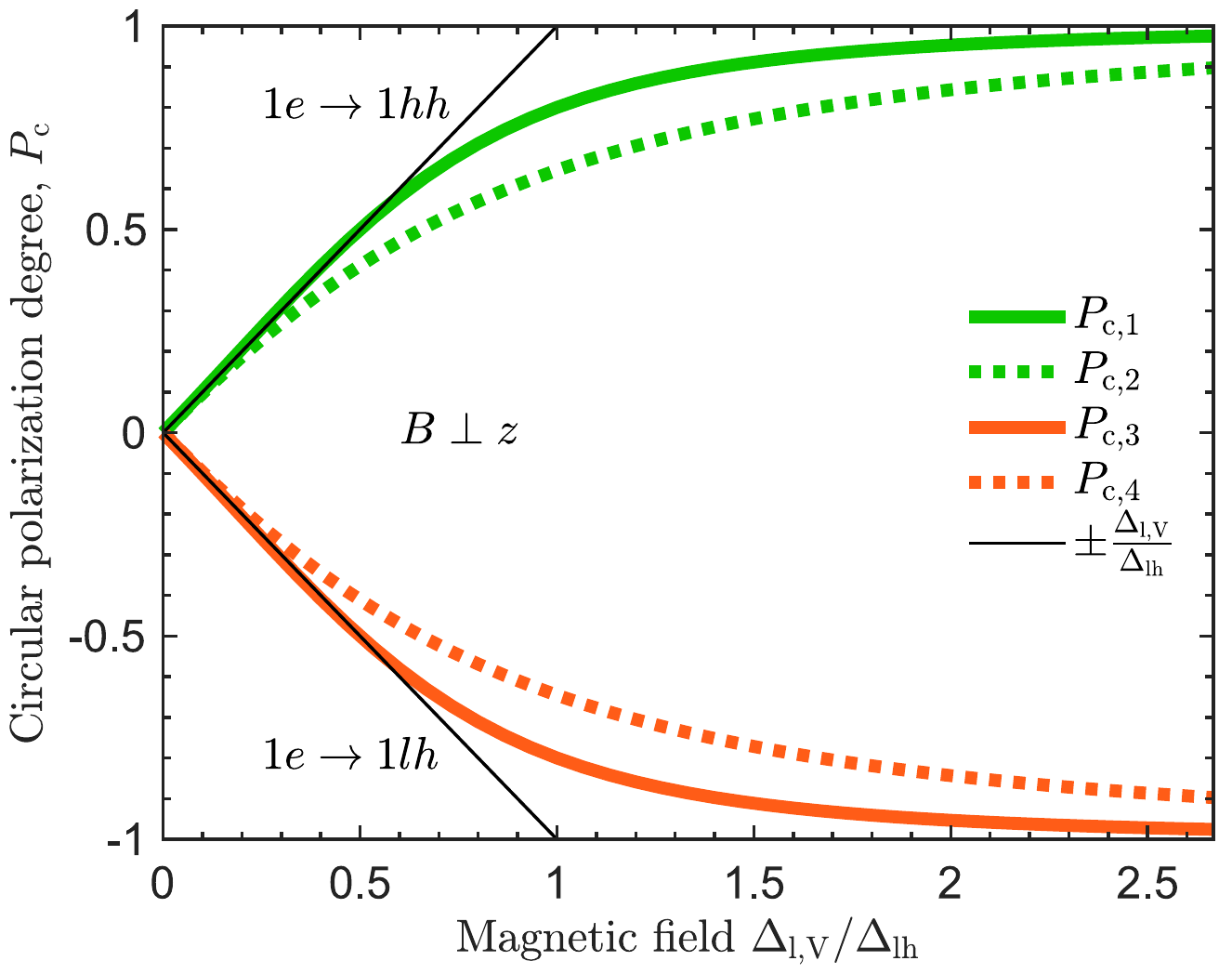}
	\caption{Magnetic field dependence of the polarization degree $P_\mathrm{c}$ for the optical transitions $1e-1hh$ (green lines) and $1e-1lh$ (orange lines), as indicated in Fig.~\ref{fig:pola1}. Bold curves (solid and dotted) have been calculated following Eqs.~\eqref{eq:Pc0}, the thin black lines correspond to the approximate expressions Eq.~\eqref{eq:Pc}.}
	\label{fig:pola2}
\end{figure}

In this work, we are more interested in the Voigt geometry, where the magnetic field lies in the plane of the quantum well ($\bm B\parallel x\perp z$) as shown in Fig.~\ref{fig:pola1}(b). For small magnetic fields, the Zeeman splitting of the electron and light-hole states is then linear in magnetic field. The Zeeman splitting of the light-holes in Voigt geometry (see Fig.~\ref{fig:pola1}(b)) is given by 
\begin{equation}
	\Delta_\mathrm{l,V}=2\Delta_\mathrm{l,F}\text{.}
	\label{eq:Zeeman_Voigt}
\end{equation}
On the other hand, the Zeeman splitting of the heavy holes in Voigt geometry is cubic in $B$ and is determined by the mixture of light- and heavy-hole states. If the electric field of the electromagnetic wave is along the external magnetic field, the optical transitions are linearly polarized, as shown by the $\pi$ arrows in Fig.~\ref{fig:pola1}(b). More importantly, the  magnetic field also induces nonzero circular polarization of the optical transitions $1,2,3$ and $4$ for the electric field in the $yz$-plane normal to the QW, that are shown by orange and green arrows in Fig.~\ref{fig:pola1}(b). It is this circular polarization that enables the TMRLE effect, i.e.\ directional emission of surface plasmons, that carry effective transverse spin along the $x$-direction, locked to the plasmon propagation direction~\cite{FelixNature-2018}. The degree of circular polarization, determining the emission directionality, is not equal to 100\% and strongly depends on the magnetic field, as shown by the calculation in Fig.~\ref{fig:pola2}. Qualitatively, the polarization degree is governed by the competition of the quantization inside the QW, that tries to pin the hole angular momentum along the $z$-axis, and the magnetic field, that tries to align the angular momentum along the $x$-axis and mixes light- and heavy-hole states. As a result, the polarization degree $P_\mathrm{c}$ is zero for $B=0$ and increases with $B$. Specifically, the polarization degrees $P_{\mathrm{c},i}$ for the optical transitions $i = 1,2,3,4$ are given by 
\cite{FelixNature-2018}
\begin{equation}
	\begin{gathered}
		P_{\mathrm{c},1}=-P_{\mathrm{c},3} = {\frac {-4\,{Z}^{2}+ 4(Z_{-}+1) Z-2\,Z_{-}+2}{4\,{Z}^{2}-4(Z_{-}+1) Z+2\,Z_{-}+7}}\text{,}\\
		P_{\mathrm{c},2}=-P_{\mathrm{c},4}= {\frac {4\,{Z}^{2}+ 4\left( Z_{+}+1 \right) Z+2\,Z_{+}-2}{4\,{Z}^{2}+ 4\left( Z_{+}+1\right) Z+2\,Z_{+}+7}}\text{,}\\
		Z=\frac{\Delta_\mathrm{l,V}}{\Delta_{lh}}\text{,}\quad Z_{\pm}=\sqrt{Z^{2}\pm Z+1}\:\text{,}\label{eq:Pc0}
	\end{gathered}
\end{equation}
where $\Delta_\mathrm{l,V}$ is the Zeeman-splitting of light holes in the Voigt geometry (see Eq.~\eqref{eq:Zeeman_Voigt}) which is the main parameter used to describe $P_{\mathrm{c},i}$ for both the heavy- and light-hole optical transitions.

Interestingly, as shown by the calculation in Fig.~\ref{fig:pola2}, the $1e-1hh$ and $1e-1lh$ transitions have pairwise opposite polarization degrees,
i.e.\ $	P_{\mathrm{c},1} = -P_{\mathrm{c},3}$ and $ P_{\mathrm{c},2}= -P_{\mathrm{c},4}$, even though the light- and heavy-hole Zeeman splittings in the Voigt geometry are very different. This is because the heavy-hole transition polarization in the Voigt geometry is determined by the light-heavy hole mixing, rather than directly by the Zeeman splitting. The polarization can be most easily  understood qualitatively in the limit of large magnetic fields, where the Zeeman splitting greatly exceeds the light-heavy hole splitting. The valence band states can then be characterized by a certain angular momentum projection $J_x=\pm 3/2,\pm 1/2$  on the field direction, the selection rules for optical transitions enforce the conservation of angular momentum along the $B$ axis and the transitions are circularly polarized,
$ P_{\mathrm{c},1} = P_{\mathrm{c},2} = \pm 1$ and $ P_{\mathrm{c},3} = P_{\mathrm{c},4} = \mp 1$. 
For small magnetic fields Eqs.~\eqref{eq:Pc0} simplify to
\begin{equation}
	P_{\mathrm{c},1/2} = -P_{\mathrm{c},3/4} \approx \pm \frac{\Delta_\mathrm{l,V}}{\Delta_\mathrm{lh}}\text{,}
	\label{eq:Pc}
\end{equation}
so that the degree of circular polarization in the Voigt geometry grows linearly in magnetic field.
Further, it is independent of the electron Zeeman-splitting and only dependent on the Zeeman splitting of the light hole.
Equation~\eqref{eq:Pc} is valid in the limit of small magnetic fields, i.e.\ $P_\mathrm{c} \ll 1$ due to $\Delta_\mathrm{l,V} \ll \Delta_{\rm lh}$. We stress that the circular polarization for both heavy-hole optical transitions $1$ and $2$ is the same in the limit of small magnetic fields, and is linear in magnetic field. Hence, even though in the Voigt geometry the Zeeman splitting of the two heavy-hole states is small so that their populations  are the same, the heavy-hole optical transitions can still have significant transverse circular polarization, enabling the TMRLE effect.

We note that at low temperatures, below the binding energy of excitons, the PL is given by the exciton emission rather than the radiative recombination of photoexcited electrons and holes discussed so far.
In our case the selection rules for the optical transitions and their polarization are the same in both descriptions (while neglecting e.g.\ the exciton fine structure) so that both descriptions are assumed to be equal for TMRLE hereafter.

\section{Experimental}
\label{sec:Experimental}
\subsection{Hybrid plasmonic-semiconductor QW structure}

\begin{figure}
	\includegraphics[width=0.45\textwidth]{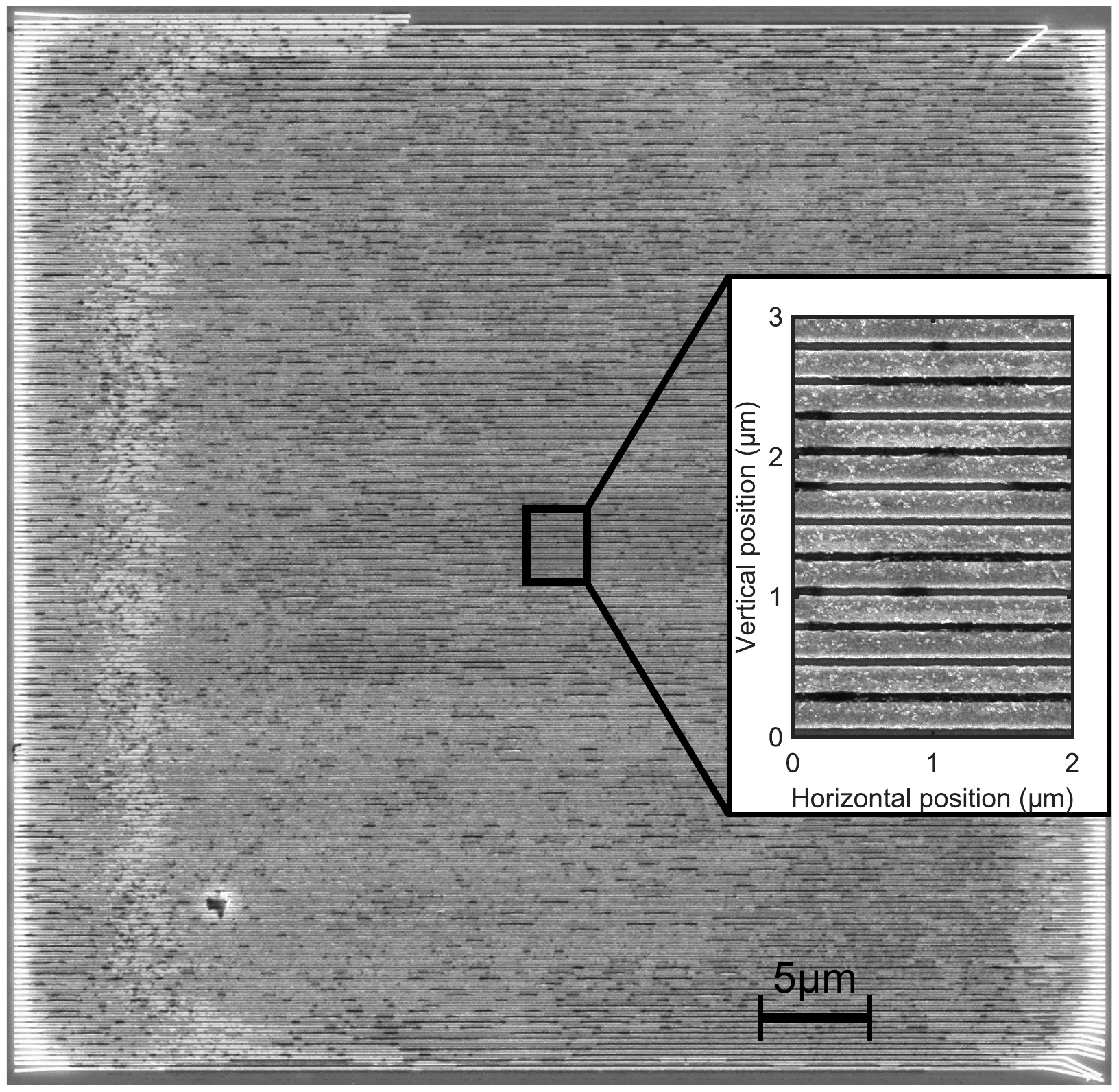}
	\caption{Scanning electron microscope (SEM) image of the $50\times50\,\si{\micro\metre\squared}$ gold grating and exemplarily a close-up from the center of the grating.
	The period and slit width correspond to $\SI{250}{nm}$ and $\SI{55}{nm}$, respectively.}
	\label{fig:SEM}
\end{figure}

The investigated sample is a planar hybrid plasmonic-semiconductor QW structure (short: hybrid structure) which comprises a semiconductor QW and a one-dimensional gold grating located in direct proximity of each other. 
Similar structures were used for the demonstration of TMRLE~\cite{FelixNature-2018} and optical orientation of electrons via plasmon to exciton spin conversion~\cite{Akimov-Vondran-PRB}.

The semiconductor part (sample number $022818\mathrm{A}$) was grown using molecular beam epitaxy on a semi-insulating (100) GaAs substrate. 
As shown in Fig.~\ref{fig:1}(a) it consists of a $\SI{10}{\nano\meter}$ $\mathrm{Cd}_{0.96}\mathrm{Mn}_{0.04}\mathrm{Te}$ DMS QW sandwiched between two layers of non-magnetic Cd$_{0.75}$Mg$_{0.25}$Te with a wider bandgap, serving as potential barriers for electrons and holes. 
These two layers are a $\SI{4.6}{\micro\metre}$ thick buffer layer between the substrate and the QW and a $\SI{30}{\nano\metre}$ thin cap layer on top of the QW.
The resulting structure has a type-I band alignment where both types of carriers (electrons and holes) are confined within the DMS (Cd,Mn)Te layer. 
The Mg content of the barriers was estimated from photoluminescence measurements using the material properties in Ref.~\cite{Hartmann1996} and the Mn concentration $x$ of the QW from magneto-PL measurements (see Appendix).
The emission from the QW is centered around $\SI{1.687}{eV}$ while the barriers emit around $\SI{2.105}{eV}$.

On top of the cap layer one-dimensional gold gratings of $50\times50\,\si{\micro\metre\squared}$ area were patterned using electron beam lithography and subsequent lift-off processing. 
The details of the patterning were reported in Ref.~\onlinecite{FelixNature-2018}. 
The gratings have a thickness of about $\SI{45}{nm}$ with a period and slit width of $\SI{250}{\nano\metre}$ and $\SI{55}{nm}$, respectively.
The period is accurate due to the high precision of the electron beam lithography technique. 
The slit width depends on the subsequent lift-off processing and was determined from scanning electron microscopy (SEM) measurements. 
A SEM image of the examined gold grating is shown in Fig.~\ref{fig:SEM} along with a close-up from the center of the grating.
Notably there are no missing gold stripes in the main area of the grating.
This demonstrates the high quality of the patterned structure despite the absence of an adhesion layer at the interface between semiconductor and gold. 
Large areas around the plasmonic gratings are left uncovered which allows us to compare TMRLE between the bare semiconductor and the hybrid structure. 
This bare part of the structure is called "bare QW" in the following chapters.

\subsection{Experimental setup and evaluation of TMRLE magnitude}

The sample is mounted on the cold finger of a liquid helium flow cryostat with variable flow rate. 
A temperature sensor close to the heat exchanger in the cryostat allows us to monitor the temperature. 
A heating element in combination with a proportional–integral–derivative (PID) control circuit can be used to heat the sample to temperatures of about $\SI{50}{K}$. 
External magnetic fields of up to $\SI{500}{mT}$ are applied in $x$-direction (parallel to the gold grating stripes) using a resistive electromagnet.

\begin{figure*}
	\includegraphics{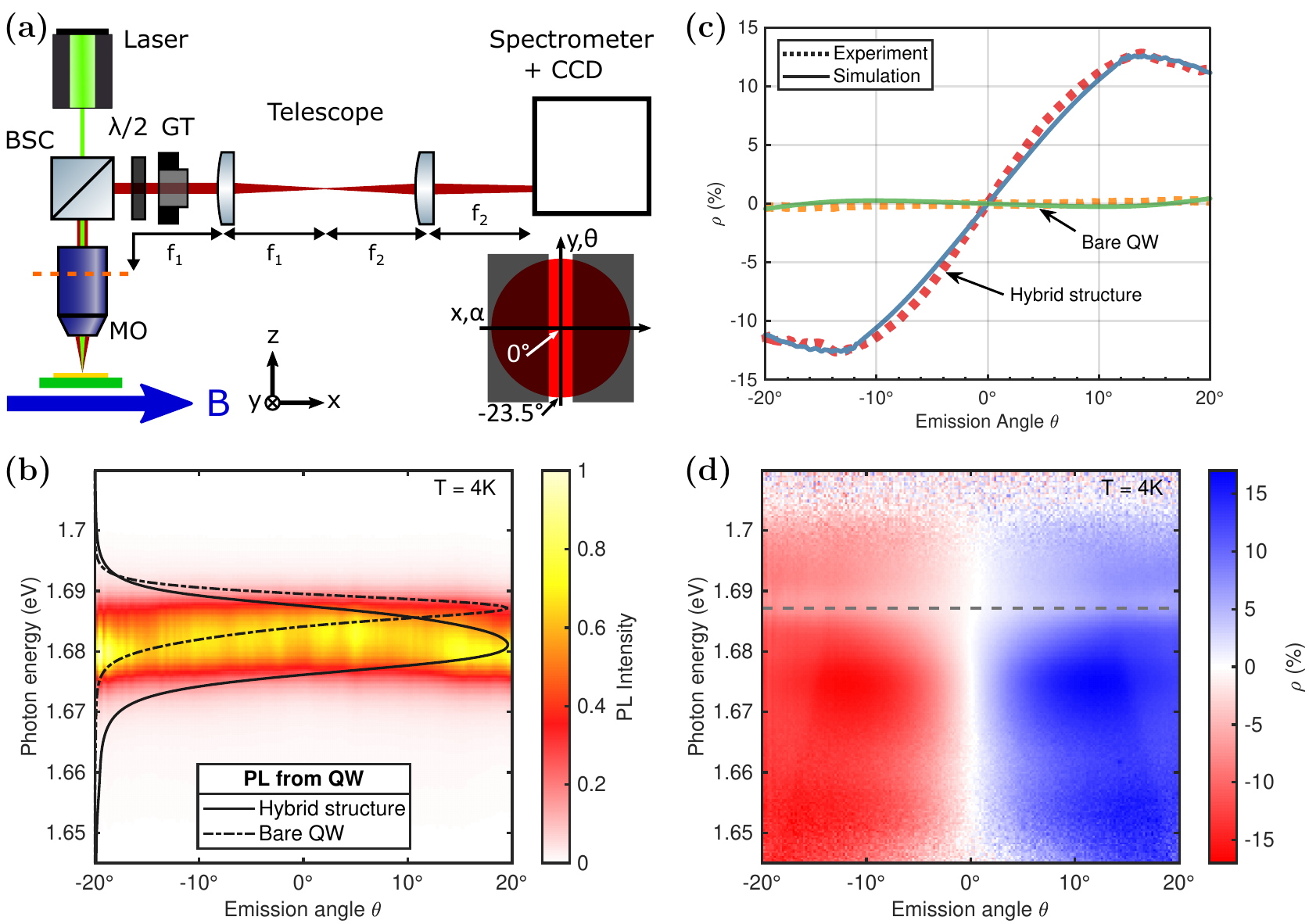}
	\caption{(a) Schematic presentation of the Fourier imaging setup used to measure the angular and spectral dependence of the light emitted from the structure. 
	MO is the microscope objective, BSC the beamsplitter cube, $\lambda/2$ and GT a half-wave plate and Glan-Thompson prism. 
	The two lenses (with focal lengths $f_{1} = \SI{400}{mm}$ and $f_{2} = \SI{300}{nm}$) form a telescope and project the Fourier plane (orange dashed line) onto the spectrometer slit.
	The magnetic field $B$ is applied in $x$-direction.
	Bottom right: Schematic presentation of the Fourier plane (red) mapped onto the vertical spectrometer entrance slit.
	(b) Two-dimensional plot showing the PL intensity distribution from the quantum well (QW) in the hybrid structure as a function of photon energy $E$ and angle of emission $\theta$ at $T = \SI{4}{K}$. 
	In addition, normalized intensity spectra for the hybrid structure and the bare QW are shown as solid and dash-dotted lines, respectively, each integrated over all angles of light emission between $\theta = \SI{\pm20}{\degree}$. 
	(c), (d) Relative change of PL intensity $\rho$ induced by the magnetic field (see Eq.~\eqref{eq:rho}) at $T = \SI{4}{K}$: 
	(c) Angular dependence $\rho(\theta)$ at the PL maxima of $\SI{1.681}{eV}$ (hybrid structure) and $\SI{1.687}{\eV}$ (bare QW), comparing experimental data at $B=\SI{485}{mT}$ (dotted lines) and simulations (solid lines).
	(d) Two-dimensional plot of $\rho(E,\theta)$ for the hybrid structure at $B=\SI{485}{mT}$, showing $\rho$ of up to $\SI{15}{\percent}$. Horizontal dashed line corresponds to the maximum of PL intensity from the bare QW structure (see panel (b)).
	}
	\label{fig:5}
\end{figure*}

To measure the TMRLE a Fourier imaging setup~\cite{Richard2005} is used, as depicted in Fig.~\ref{fig:5}(a). 
The sample is excited off-resonantly with a $\SI{552}{\nm}$ ($\SI{2.25}{eV}$) solid-state laser in continuous wave mode, which corresponds to above barrier excitation. 
The laser light is focused onto the sample using a $20\times$ microscope objective (MO) with a numerical aperture of $\num{0.4}$. 
The diameter of the spot is about $\SI{7}{\micro\m}$ and the corresponding power density in all measurements is kept at about $\SI{400}{\watt\per\centi\m\squared}$ unless stated otherwise. 
The light emitted from the sample is collected by the same microscope objective in back scattering geometry, leading to a measurable range of emission angles between $\SI{\pm23.5}{\degree}$. 
In a distance of twice the MO focal length from the sample the Fourier plane is located. 
Here the emission angle of the light from the sample $\theta$ is transferred into a spatial offset from the optical axis, giving direct access to the angular distribution of the emitted light. 
Using a telescope consisting of two lenses the Fourier plane is mapped onto the $\SI{150}{\micro\meter}$ wide entrance slit of an imaging single stage $\SI{0.5}{m}$ spectrometer with linear dispersion of $\SI[per-mode=symbol-or-fraction]{6.43}{\nm\per\mm}$. 
At the entrance slit the light spot contains the emission angle information in the horizontal ($\widehat{=} \alpha$) and vertical direction ($\widehat{=}\theta$), both perpendicular to the optical axis.
As shown schematically in Fig.~\ref{fig:5}(a), the vertical entrance slit cuts off most of the horizontal portion of the light spot, leaving only a vertical slice to enter the spectrometer.
The light entering the spectrometer is thus limited to the PL emitted from the sample at around $\alpha = \SI{0\pm0.8}{\degree}$ in horizontal direction and $\theta$ between $\SI{\pm23.5}{\degree}$ (vertical direction). 
The spectrometer horizontally separates the light into its spectral components, leaving the angular information in the vertical direction ($\theta$) intact.
This angle- and spectrally- resolved PL signal $I(E,\theta)$ is then detected by a charge coupled device (CCD) camera, leading to a spectral resolution of $\SI{1}{nm}$.
Behind the MO the combination of a half-wave plate and a Glan-Thompson prism is used to select the desired linear polarization of the emitted light. 
Since the TMRLE occurs only for p-polarized emission~\cite{FelixNature-2018}, all data shown here were measured in p-polarization. 
A longpass filter removes residual laser light in the detection path. 

The magnitude of TMRLE is determined by the degree of directionality $C$, which is defined as relative difference in the emission intensities for positive and negative emission angles $I(+\theta)$ and  $I(-\theta)$, respectively, 
\begin{equation}
	C=\frac{I(+\theta)-I(-\theta)}{I(+\theta)+I(-\theta)}.
	\label{eq:C-def}
\end{equation}
It can be evaluated from the magnetic-field-induced changes in the angle-resolved PL data. 
Just like the transverse magneto-optical Kerr effect (TMOKE), the TMRLE magnitude is an odd function of in-plane magnetic field. 
Therefore, it is convenient to analyze the difference of the measured light intensities $I$ for opposite directions of the magnetic field $\pm B$. 
The relative change of intensity induced by the magnetic field is determined as
\begin{equation}
	\rho = \frac{I(+B) - I(-B)}{I(+B) + I(-B)}.
	\label{eq:rho}
\end{equation}
Then the magnetic-field-induced directionality can be expressed as antisymmetric part of $\rho$ with respect to the emission angle
\begin{equation}
	C(\theta) = [\rho(\theta) - \rho(-\theta)]/2.
	\label{eq:C-via-rho}
\end{equation}
Hence, by taking angle- and spectrally-resolved PL intensity distributions for two opposite magnetic field directions, the parameters $\rho$ and $C$ characterizing TMRLE are deduced. 
The noise level can be reduced by repetitive switching of the magnetic field direction and subsequent averaging.
Typically for one measurement the magnetic field direction was switched back and forth $25$ times and each time $16$ spectra were taken with \SI{3}{s} exposure time for each spectrum.

\section{TMRLE in ${\rm \bf (Cd,Mn)Te}$ based structure}
\label{sec:TMRLE-DMS}
\subsection{Routing of PL at low temperatures}
\label{subsec:lowT-TMRLE}
Figure~\ref{fig:5}(b) exemplarily shows a two-dimensional plot of the angle- and spectrally-resolved intensity distribution $I(E,\theta)$ of the PL from the QW in the hybrid structure. 
The intensity shown is the average of the intensities for positive and negative magnetic field direction $\pm B$. 
The data are taken at low temperature $T=\SI{4}{K}$ (measured at the cryostat temperature sensor) and the strength of magnetic field corresponds to $B = \SI{\pm485}{mT}$.
The PL is centered around $\SI{1.681}{eV}$ and shows a weak angular dependence where the PL maximum shifts to lower energies with increasing emission angle. 
The intensity spectrum of the bare QW (without gold grating on top) does not show this curvature and differs in spectral position and width.
This comparison is also shown in Fig.~\ref{fig:5}(b) as overlay, where the normalized, angularly integrated PL intensity spectra for both the hybrid structure with gold grating and the bare QW are displayed as solid and dash-dotted lines, respectively.
The PL from the bare QW has its maximum at $\SI{1.687}{eV}$ with a FWHM of about $\SI{5.4}{meV}$, whereas the hybrid structure has a larger FWHM of about $\SI{11}{meV}$ with its maximum located at the slightly lower energy of $\SI{1.681}{eV}$. 
The shift towards lower PL energies in the hybrid structure is attributed to the formation of a Schottky barrier at the metal-semiconductor interface, which bends the band structure and lowers the energy of interband optical transitions in the QW due to the Stark effect. 
The increase in spectral width could originate from the inhomogeneous distribution of the electric field in the plane of the structure due to different band bending under metal stripes and slits. 

By comparing the measured light intensities $I$ for opposite magnetic field directions $\pm B$ according to Eq.~\eqref{eq:rho} the parameter $\rho$ is obtained, quantifying PL intensity changes induced by the magnetic field.
The resulting angular dependence $\rho(\theta)$ at $T = \SI{4}{K}$ is shown in Fig.~\ref{fig:5}(c), again comparing the hybrid structure (red) and the bare QW (orange). 
The intensity spectra were averaged along the energy axis in a $\SI{10}{\milli\eV}$ window centered around their respective PL maximum, before calculating $\rho$ according to Eq.~\eqref{eq:rho}.
For the hybrid structure we observe $\rho(\theta)$ rising monotonously from $\SI{0}{\percent}$ at $\theta=\SI{0}{\degree}$ towards maximum and minimum values of $\SI{\pm13}{\percent}$ at $\theta=\SI{\pm14}{\degree}$ and decreasing slowly for larger emission angles. 
The effect is hardly seen on the bare QW structure due to the absence of true surface waves, e.g.\ SPPs, in the vicinity of the QW layer. Weak directional emission $\rho<\SI{0.5}{\percent}$ takes place though because of a far-field routing effect which is governed by the interference of directly emitted light beams and those reflected at the (Cd,Mg)Te/GaAs buffer-substrate interface~\cite{FelixNature-2018}.

The two-dimensional angular and spectral dependence of $\rho(E,\theta)$ for the hybrid structure is shown as color plot in Fig.~\ref{fig:5}(d), again for $T = \SI{4}{K}$. 
The colors blue and red represent different signs of $\rho$ and their saturation visualizes the magnitude. 
Hence, blue colors represent more light emitted for positive than for negative magnetic fields at that angle and photon energy and red vice versa. 
The spectral dependence of $\rho$ is fairly flat, reaching up to $\rho = \SI{15}{\percent}$ around $\SI{1.675}{eV}$ with a small oscillatory behavior along the energy axis due to the far-field interference effect mentioned above. 
Around $\SI{1.688}{eV}$ an area with less directional emission is visible as less saturated colors (see horizontal dashed line in Fig.~\ref{fig:5}(d)). 
This dip spectrally coincides with the PL measured from the bare QW structure. 
At first glance this feature could originate due to emission from areas with defects or missing gold stripes in the hybrid structure. However this hypothesis is excluded because of the SEM image in Fig.~\ref{fig:SEM}.
Thus we attribute this feature to weak spurious signal from outside the grating.

In agreement with previous studies in Ref.~\onlinecite{FelixNature-2018} it follows from Figs.~\ref{fig:5}(c) and \ref{fig:5}(d) that the magnetic-field-induced variation of PL intensity $\rho$ is an odd function of the emission angle $\theta$, i.e.\ $\rho(-\theta) = -\rho(\theta)$. 
Such antisymmetric behavior also holds true for all measurements presented in this work so that in the investigated structures the magnetic-field-induced changes of the PL intensity are due to the directionality effect only and $C(\theta) = \rho(\theta)$. 
Interestingly, there is an optimum angle $\theta_{\rm max}\approx 14^\circ$ in Fig.~\ref{fig:5}(c) where we observe the maximum degree of directionality. 
This angle corresponds approximately to the resonance condition when the SPP energy fits the energy of the exciton resonance $E_{\rm X}$. 
It is given by $\sin{\theta} = k_{\rm SPP}/k_0$ with the wavevector of the emitted light $k_0=E_{\rm X}/\hbar c$ and $\hbar$ the reduced Planck constant. 

The data are in good agreement with calculations based on the scattering matrix method~\cite{Whittaker1999}, which was presented in Ref.~\onlinecite{FelixNature-2018} for a similar structure. 
The calculations of the PL spectra model the excitons as uncorrelated point dipoles with fixed polarization based on the external magnetic field direction, randomly distributed in a distance $d = \SI{30}{nm}$ from the plasmonic interface. 
The results of the calculations are shown in Fig.~\ref{fig:5}(c) by solid lines both for the hybrid structure (blue) and the bare QW (green), showing good agreement with the experimental data (dotted red and orange lines, respectively).
The shown angular dependences $\rho(\theta)$ resulting from the simulations are obtained similar to their experimental counterparts by first integrating the intensity spectra along the energy axis in the same $\SI{10}{\milli\eV}$ window as the experimental data and then calculating $\rho$ according to Eq.~\eqref{eq:rho}.
For the spectral profile of the emission a Gaussian distribution is assumed with the FWHM and maximum energy taken from the experimental data in Fig.~\ref{fig:5}(b).
The energy-dependent refractive indices for gold, (Cd,Mg)Te and GaAs were taken from Refs.~\onlinecite{Johnson1972-n_Au, Andre1997-n_CdMgTe, Aspnes1986-n_GaAs}. 
The background refractive index contrast between the QW and the (Cd,Mg)Te spacer layer was neglected.

The degree of circular polarization $P_\mathrm{c} = \SI{9}{\percent}$ was used as fitting parameter to match the magnitude of directionality from the experiment on the hybrid structure. 
Following Eq.~\eqref{eq:Pc} this amounts to a Zeeman splitting of light holes in Voigt geometry of $\Delta_\mathrm{l,V} = \SI{1.8}{\milli\eV}$, using an energy splitting $\Delta_{\mathrm{lh}} \approx \SI{20}{meV}$ between heavy- and light-hole states which has been previously determined in similar structures from reflectivity spectra~\cite{Borovkova-2019}. 
The value of $\Delta_\mathrm{l,V}$ is in agreement with the expected light hole splitting at $B=485$~mT if we take into account that the Mn-spin temperature is larger than $T=\SI{4}{K}$ measured by the temperature sensor.
For low amounts of Mn-ions $x$ the Mn-spin temperature can be increased significantly by continuous-wave photoexcitation due to slow spin-lattice relaxation~\cite[Chapter~8]{Kossut-book} and corresponds to $\approx \SI{23}{K}$ in our experiments as will be shown in the following section.

Thus far the data at low temperatures clearly demonstrate that the magnetic-field-induced directionality in hybrid structures results from the generation of SPPs via exciton emission which is also in agreement with calculations based on the scattering matrix method. 
In what follows we concentrate on the temperature dependence of TMRLE in (Cd,Mn)Te QW based structures.

\subsection{Temperature dependence}
\label{sec:T-dep}

\begin{figure}
	\includegraphics{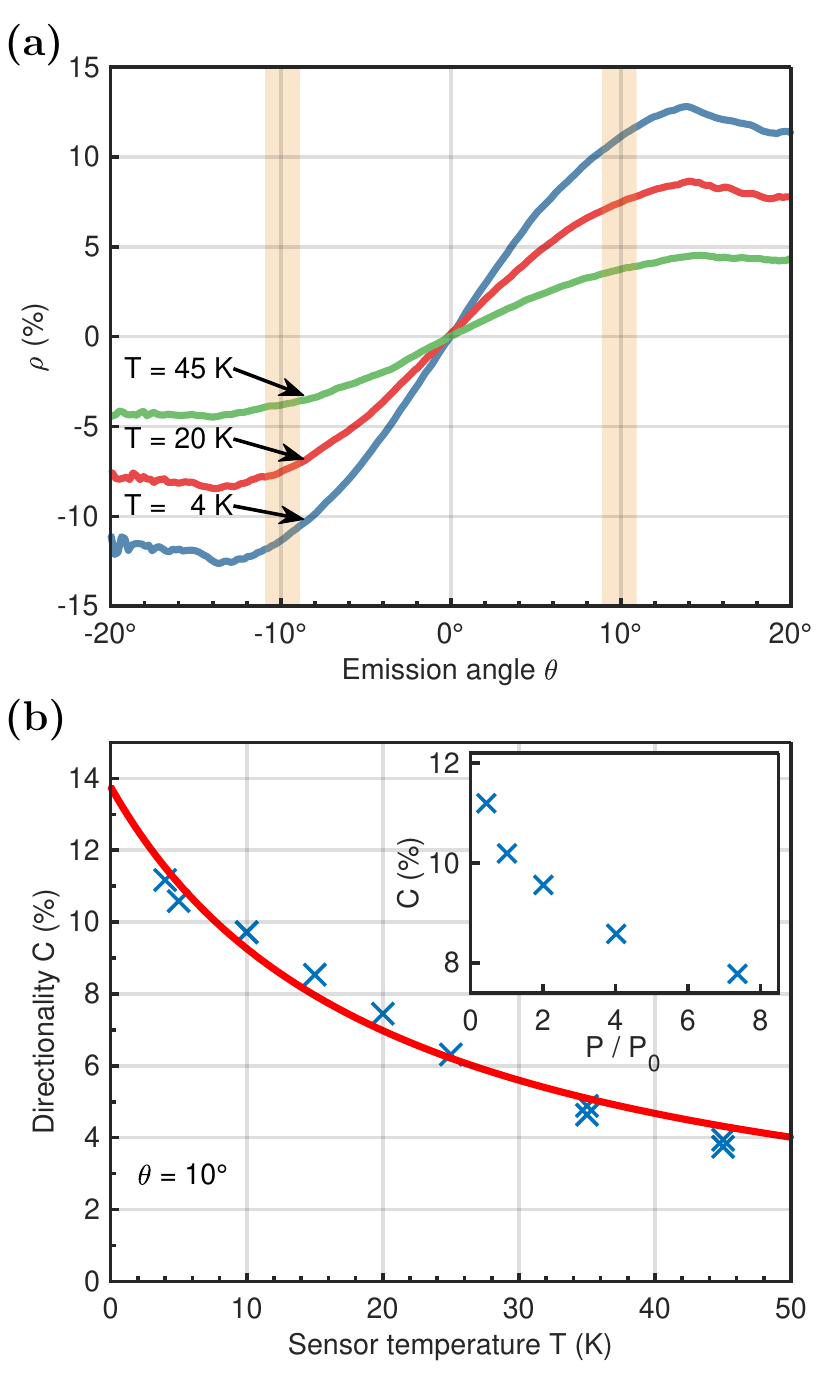}
	\caption{
		(a) Angular dependence $\rho(\theta)$ for PL from the QW in the hybrid structure for different temperatures $T$.
		For this the measured light intensities were first integrated in a $\SI{10}{\milli\eV}$ window centered around the PL maximum at $E = \SI{1.681}{eV}$. 
		The magnitude of directional emission decreases with increasing temperature while retaining the same angular dependence.
		(b) Temperature dependence $C(T)$ for the PL collected in the window of $\theta = \SIrange{9}{11}{\degree}$ [highlighted areas in panel (a)] and in the same photon energy range as above.
		The red line shows the fit using a modified Brillouin function according to Eq.~\eqref{eq:FitFunc} with $T_{\mathrm{off}} = \SI{19\pm3}{\kelvin}$.
		Inset: Dependence of $C$ on the excitation power density $P/P_0$ at $T = \SI{4}{K}$ for the same angles and photon energy with $P_0 \approx \SI{400}{\watt\per\cm\squared}$.
		Higher power densities lead to local heating and, correspondingly, decreasing PL directionality.}
	\label{fig:Tdep}
\end{figure}

The temperature dependence of the TMRLE for the PL from the QW in the hybrid structure was measured in the range from $T=\SI{4}{K}$ to $\SI{45}{K}$ for $B=\SI{485}{mT}$. 
The data are summarized in Fig.~\ref{fig:Tdep}. 
Exemplarily the angular dependence $\rho(\theta)$ for three different temperatures $T$ of $4$, $20$ and $\SI{45}{K}$ are shown in Fig.~\ref{fig:Tdep}(a). 
The PL signals are analyzed within the photon energy window of $\SI{10}{\milli\eV}$ centered around the PL maximum $E = \SI{1.681}{eV}$ in the same way as in Fig.~\ref{fig:5}(c). 
The angular dependence of $\rho(\theta)$ is the same for all measured temperatures but its magnitude decreases with increasing temperature.
Figure~\ref{fig:Tdep}(b) shows the temperature dependence of the directionality parameter $C$ which is calculated using Eq.~\eqref{eq:C-via-rho} in the range of emission angles between $9^\circ$ and $11^\circ$ [highlighted areas in panel Fig.~\ref{fig:Tdep}(a)]. 
It follows that the emission routing decreases monotonously with increasing temperature.

In (Cd,Mn)Te QW based structures the decrease of the TMRLE magnitude for larger temperature is expected due to the decrease of the magnetic susceptibility, which in turn reduces the magnitude of Zeeman splitting between the exciton spin states. 
For a transverse magnetic field, which is directed along the $x$-axis, this influences the degree of circular polarization $P_\mathrm{c}$ in the $yz$-plane for optical transitions. 
The latter directly determines the directionality parameter $C \propto P_\mathrm{c}$. 
In small magnetic fields $P_\mathrm{c} \approx \Delta_\mathrm{l,V}/\Delta_{\mathrm{lh}}$ holds true, see Eq.~\eqref{eq:Pc} above.
In order to describe the experimental data we have  determined  the effective temperature
$T_{0} = \SI{1.9}{K}$ and effective spin $S_{\mathrm{eff}} \approx \num{1.47}$  from magneto-PL measurements at $T = \SI{1.5}{K}$ in Faraday geometry (see Appendix for details). This corresponds to $x \approx \SI{4}{\percent}$, see Ref.~\onlinecite{Ossau-1993}. 

As follows from Eqs.~\eqref{eq:Pc}, \eqref{eq:Zeeman}, \eqref{eq:Brillouin}, \eqref{eq:Zeeman_Voigt} the temperature dependence of the directionality can thus be approximated as
\begin{equation}
	C(T) \propto B_{5/2}\left( \frac{5}{2} \frac{\mu_\mathrm{B} g_\mathrm{Mn} B}{k_\mathrm{B}(T_\mathrm{Mn}+T_0) } \right)\text{.}
	\label{eq:FitFunc}
\end{equation} 
In our experiment the temperature $T$ measured by the sensor is not identical to the Mn-spin temperature $T_\mathrm{Mn}$ due to several reasons.
First and most importantly, for low amounts of Mn-ions $x$ the Mn-spin temperature can be increased significantly by continuous-wave photoexcitation due to slow spin-lattice relaxation~\cite[Chapter~8]{Kossut-book}.
In addition, the sample is located on a cold finger in the cryostat at some distance from the heat exchanger where the temperature is measured, which could lead to a further, small offset.
As a result for these measurements the temperature in Eq.~\eqref{eq:FitFunc} is given by $T + T_{\rm off}+ T_0$, where $T_{\rm off}$ is the temperature difference between $T$ measured by the sensor and the Mn-spin temperature $T_\mathrm{Mn}$. 

With $T_{\rm off}$ and an arbitrary amplitude as the only fitting parameters we use Eq.~\eqref{eq:FitFunc} to describe the temperature dependence of $C(T)$ in Fig.~\ref{fig:Tdep}(b). 
The experimental data are in good agreement with the fit shown as red curve, yielding $T_{\rm off} = \SI{19\pm3}{K}$ (standard error).
It should be noted that in our model we assume $T_\mathrm{off}$ to be constant for all temperatures $T$, even though a small dependence can be expected because an increasing (lattice) temperature leads to a more effective cooling of the Mn-system via spin-lattice relaxation~\cite[Chapter~8]{Kossut-book}, possibly explaining some deviations in Fig.~\ref{fig:Tdep}(b).
The notably large offset between the sensor temperature and the Mn-spin temperature is explained by the relatively high laser power density $P_0 \approx \SI{400}{\watt\per\cm\squared}$ heating the Mn system.
The heating due to laser excitation is confirmed by evaluating the directionality $C$ at $T = \SI{4}{\kelvin}$ for different laser power densities. 
As shown in the inset of Fig.~\ref{fig:Tdep}(b) the average directionality $C$ at $\theta = \SIrange[range-phrase= -]{9}{11}{\degree}$ and $E = \SI{1.681}{eV}$ decreases with increasing power density $P$. 
In addition it should be noted that recent studies of spin dynamics in similar hybrid structures pointed out the importance of local heating effects in the presence of SPPs~\cite{Akimov-Vondran-PRB}.

Thus far we have shown that the magnitude of directional light emission due to TMRLE in DMS (Cd,Mn)Te based structures is strongly influenced by the temperature. 
Evaluating Eq.~\eqref{eq:FitFunc} at $\SI{200}{K}$ and $\SI{300}{K}$ based on the experimental data from Fig.~\ref{fig:Tdep}(b) (and for an increased magnetic field $B = \SI{1}{T}$) yields $C = \SI{2.6}{\percent}$ and $\SI{1.8}{\percent}$, respectively. 
These values represent lower limit estimates though and should be even larger if the intrinsic g-factor of the holes $g_\mathrm{h}\approx\num{-0.5}$ in CdTe is taken into account~\cite{Debus2013, Bartsch2011}. This contribution has the same sign as the exchange term and, therefore, leads to an increase of the Zeeman splitting (see Eq.~\eqref{eq:Zeeman}).

\subsection{The role of light-hole excitons}

\begin{figure}
	\includegraphics{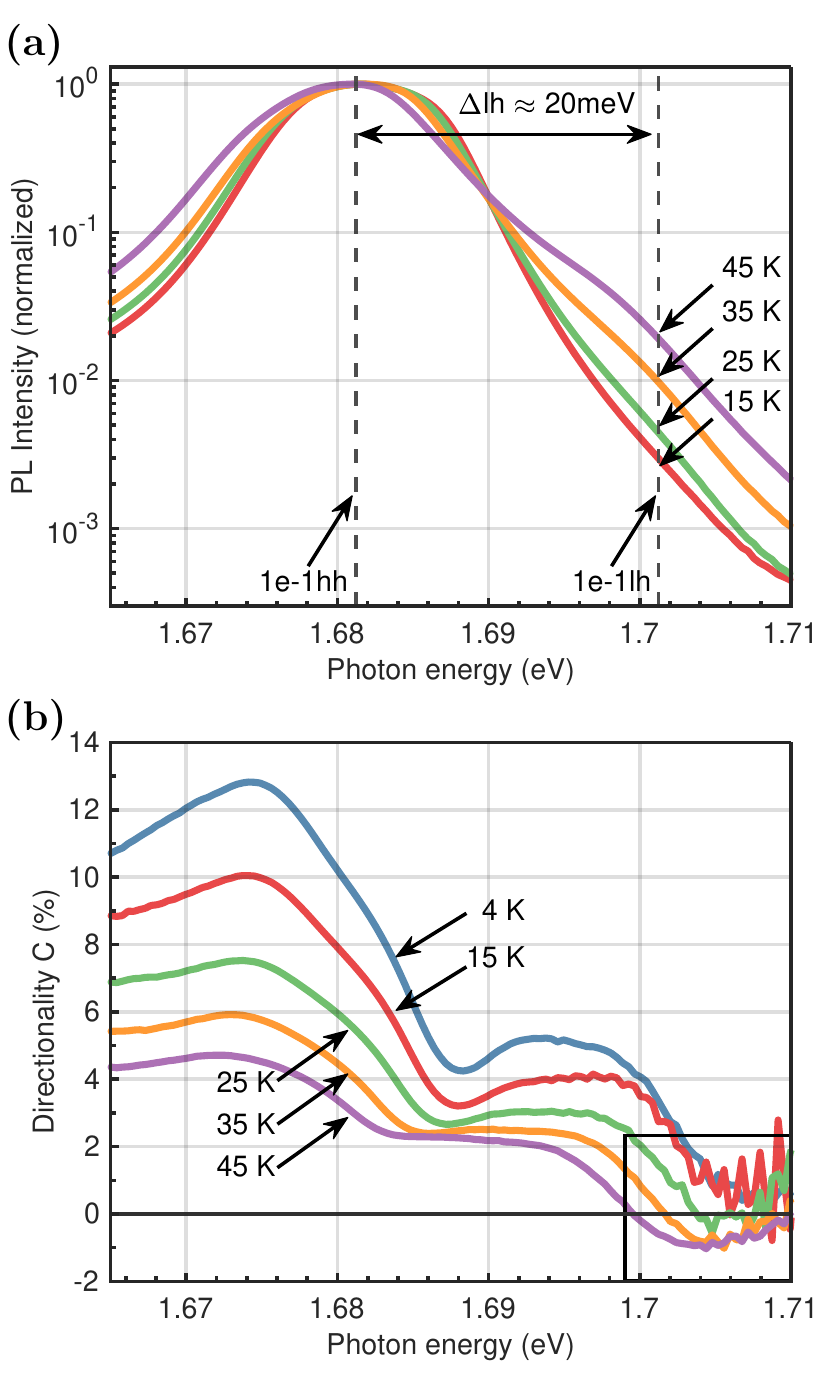}
	\caption{
		(a) Normalized PL spectra from the hybrid structure in logarithmic scale for different temperatures $T$. 
		The data are averaged over emission angles $\theta = \SI{-20}{\degree}$ to $\SI{20}{\degree}$.
		For $T = \SI{35}{K}$ and $\SI{45}{K}$ the PL from light-hole excitons becomes visible around $\SI{1.7}{eV}$, matching the energy splitting $\Delta_{\mathrm{lh}} \approx \SI{20}{meV}$ between heavy- and light-hole states. 
		(b) Spectral dependence of the directionality $C(E)$ for the hybrid structure. 
		For this the PL intensity was first averaged separately for negative ($\SI{0}{\degree}$ to $\SI{-20}{\degree}$) and positive angles ($\SI{0}{\degree}$ to $\SI{+20}{\degree}$), then $\rho(E)$ was calculated for each range using Eq.~\eqref{eq:rho} and finally $C(E)$ by using Eq.~\eqref{eq:C-via-rho}. 
		For $\SI{35}{K}$ and $\SI{45}{K}$ a sign change of the directionality $C$ appears at photon energies above $\SI{1.7}{eV}$ (see black square) due to appearance of light hole emission with opposite circular polarization in the $yz$-plane compared to the heavy-hole emission.}
	\label{fig:lh}
\end{figure}

The directionality due to the TMRLE examined so far has its origin in the PL from the QW due to the recombination of heavy-hole excitons. 
For small magnetic fields (i.e. $\Delta_\mathrm{l,V}\ll \Delta_{\rm lh}$) both heavy-hole exciton transitions ($1e-1hh$ transitions) correspond to the lowest exciton energy states and both have the same helicity of circular polarization $P_{\mathrm{c},i}$, $i=1,2$, in the $yz$-plane, leading to the same direction of emission for the TMRLE (see Fig.~\ref{fig:pola1}(b)). 
The light-hole ($1e-1lh$) excitons are higher energy states and their occupation is almost zero at low lattice temperatures $T_\mathrm{c}$ (i.e.\ $k_{\rm B}T_\mathrm{c} \ll \Delta_{\rm lh}$) due to fast energy relaxation into the lowest energy heavy-hole states within the exciton lifetime. 
The population of light holes $\propto \exp\left(- \Delta_\mathrm{lh}/k_\mathrm{B}T_\mathrm{c}\right)$ rises with increasing temperature and the recombination of light hole excitons starts to contribute to the PL signal~\cite{Weisbuch-81}. 
As discussed in Section~\ref{sec:Polarization}, the $1e-1lh$ optical transitions have the opposite circular polarization in the $yz$-plane compared to the heavy-hole transitions, leading to their emission routed into the opposite direction and thus to the opposite directionality $C$. 
Therefore, the contribution from light holes at elevated temperatures is expected to play an important role in the routing of PL from the QW.

Figure~\ref{fig:lh}(a) shows the normalized spectral dependence of the emitted light intensity in logarithmic scale for different temperatures. 
These data are extracted from the same measurements as before, now by integrating over all angles of light emission between $\SI{\pm20}{\degree}$. 
For the higher temperatures of \SIlist{25;35;45}{\kelvin} an emission shoulder around $\SI{1.7}{eV}$ becomes visible in the logarithmic scale. 
This emission peak originates from the recombination of light-hole excitons which is separated from the heavy-hole excitons at $\SI{1.681}{eV}$ by the $1lh-1hh$ splitting of $\Delta_\mathrm{lh} \approx \SI{20}{meV}$. 
The relative intensity of the high energy signal compared to the main peak of the heavy-hole exciton increases for higher temperatures, due to the occupation by thermal activation as mentioned above.

The directionality $C(E)$ is shown in Fig.~\ref{fig:lh}(b) in the same spectral range for five exemplary temperatures. 
It has its maximum on the lower energy side of the heavy-hole $1e-1hh$ optical transition. 
In this range the directionality decreases but its spectral dependence remains the same. 
At larger photon energy around 1.688~eV at $T=4$~K there is a dip which coincides with the PL maximum from the bare QW as discussed in Section \ref{subsec:lowT-TMRLE}. 
The position of the dip shifts to lower energies with increasing temperature, which is due to the reduction of the energy band gap at larger temperatures~\cite{Hwang2012-Eg-CdTe}. 
This is also in agreement with the red shift of the PL maximum in Fig.~\ref{fig:lh}(b). 
The most striking behavior occurs at even larger photon energies in the vicinity of the light-hole exciton emission at $\numrange[]{1.7}{1.71}\,\si{eV}$. 
Here, $C$ is close to $0$ for low temperatures. For $\SI{35}{K}$ and $\SI{45}{K}$ however, the directionality becomes negative in this spectral range, which is in full accord with our expectations for the light-hole exciton contribution presented in Section \ref{sec:Polarization}.

Therefore, we demonstrate that the emission related to the recombination of light holes indeed shows opposite directionality as compared to heavy holes and its contribution becomes important at higher temperatures. 
In this case the routing is diminished as soon as the thermal energy becomes significantly larger than the splitting $\Delta_{\rm lh}$, unless spectral filtering between the heavy- and light-hole PL bands is possible. 
Moreover, it should be noted that $C\propto 1/\Delta_{\rm lh}$ and larger values of $\Delta_{\rm lh}$ reduce the overall magnitude of TMRLE.

\section{TMRLE in non-magnetic QW structures}
\label{sec:highT}
In structures based on the DMS (Cd,Mn)Te we observe a decrease of the TMRLE magnitude with increasing temperature. 
This is related to the decrease of manganese spin polarization $\left<S_z^\mathrm{Mn}\right>$, i.e.\ magnetic susceptibility, which consequently reduces the magnitude of the light-hole Zeeman splitting $\Delta_{\rm l,V}$ and thus the circular polarization in the $yz$-plane $P_\mathrm{c}$ of the optical transitions. 
While ferromagnet based structures are attractive for routing applications, several difficulties such as potential low-temperature ferromagnetic phases and strong non-radiative decay of photoexcited carriers in these materials remain unresolved~\cite{Dietl-FM, Sapega-GaMnAs}.
Another way to achieve noticeable routing at larger temperatures is to use a non-magnetic semiconductor constituent (i.e. without magnetic ions) with intrinsically large hole Land\'e g-factor $g_{\rm h}$ which (following Eqs.~\eqref{eq:Zeeman} and \eqref{eq:Zeeman_Voigt}) results in $\Delta_{\rm l,V} = 2g_{\rm h} \mu_B B$, independent of the temperature.
Narrow bandgap semiconductors such as InAs posses large g-factors. 
In what follows we model a structure based on an (In,Ga)As/(In,Al)As QW system with photon energies of emission in the telecommunication spectral range (around \SI{1600}{nm} or \SI{0.78}{eV}) and a temperature dependence of directional emission very different from the structure presented before.

Exact values for the hole g-factor are not available. Therefore, we make a rough estimate of $g_{\rm h}$ based on the effective energy bandgap, in our case that is the energy of the $1e-1hh$ optical transition in the QW.
In Ref.~\onlinecite{Belykh-2016} Belykh et al.\ measured directly the g-factor of holes in $\mathrm{InAs/In}_{0.53}\mathrm{Al}_{0.24}\mathrm{Ga}_{0.23}\mathrm{As}$ self-assembled quantum dots for different energies using time-resolved ellipticity measurements in transmission geometry, allowing the estimation of $3g_\mathrm{h} \approx 4$ at $\SI{0.77}{eV}$. 
Even higher hole g-factors up to $15$ were evaluated from magneto-PL data in a $\SI{4}{nm}$ InAs QW with $\mathrm{In}_{0.75}\mathrm{Al}_{0.25}\mathrm{As}$ barriers in the photon energy range of 0.58 to 0.72~eV~\cite{Ganichev-2017}. 
We note, however, that none of these particular structures would allow one to achieve optimum conditions for routing because the lattice mismatch between the active region (QW or quantum dots) and the surrounding matrix (barriers) produces strain which increases the splitting between the heavy- and light-holes $\Delta_{\mathrm{lh}}$ and thus reduces $P_\mathrm{c}$. 
With a lower estimate of $3g_\mathrm{h} = 4$ for an (In,Ga)As/(In,Al)As QW structure we obtain a Zeeman splitting of the hole levels of $\Delta_\mathrm{l,V} = \SI{0.15}{meV}$ in a magnetic field of $\SI{1}{\tesla}$. 

\begin{figure*}
	\includegraphics[width=0.9\textwidth]{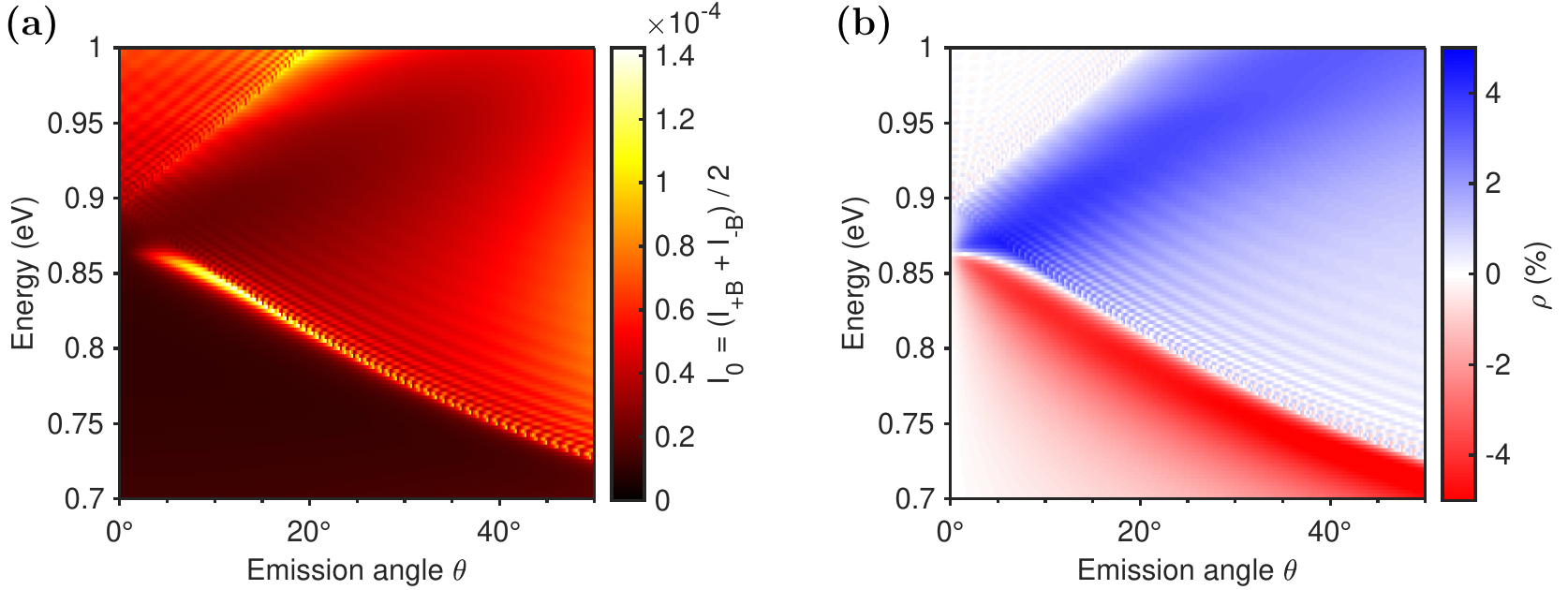}
	\caption{ 
		(a) Modeling of the spectral dependence of the intensity emitted via the plasmonic gold grating for the (In,Ga)As/(In,Al)As QW structure using the assumption that interband optical transitions contribute equally in the depicted spectral range $0.7-1.0$~eV. The intensity is calculated as an average of the two oppositely elliptically polarized emitters in the QW, i.e.\ opposite external magnetic fields. Areas of high intensity highlight the plasmonic branches of the structure. The QW is assumed to emit at all photon energies in the depicted spectral range.
		(b) Relative change of PL $\rho(E,\theta)$ with $P_\mathrm{c} = \SI{0.82}{\percent}$.}
	\label{fig:Simu}
\end{figure*}

The QW thickness $L$ as well as the compositions of both the ternary alloys used as the (In,Ga)As QW and (In,Al)As barriers in the model are optimized in order to achieve emission around $\SI{1600}{nm}$ ($\approx \SI{0.77}{eV}$) and a splitting $\Delta_{\mathrm{lh}}$ in the order of the thermal energy $k_\mathrm{B}T_\mathrm{c}$ at the desired temperature. 
Further, the lattice constants of both materials need to be matched in order to minimize the strain in the QW and thus to avoid an additional increase of $\Delta_{\mathrm{lh}}$. 
Using a one-dimensional QW potential for electrons and holes we calculated the ground state energy levels in the conduction band ($1e$) and valence bands ($1hh$ and $1lh$) in a QW with variable thickness. 
Choosing the composition of $\mathrm{In}_{0.61}\mathrm{Ga}_{0.39}\mathrm{As/}$ $\mathrm{In}_{0.60}\mathrm{Al}_{0.40}\mathrm{As}$ with an $L = \SI{12}{nm}$ thick QW leads to emission around $\SI{0.78}{eV}$ and a heavy-hole light-hole splitting of $\Delta_{\mathrm{lh}} \approx \SI{19}{meV}$, which is larger than the thermal energy $\SI{17}{meV}$ at $\SI{200}{K}$. 
In this composition the two materials are lattice matched, minimizing strain in the QW. 

The calculations were done with material properties taken from Ref.~\onlinecite{Vurgaftman2001}, using the recommended values therein, as presented in Tab.~\ref{tab:materials-parameters}.
For the constituents InAs, GaAs, AlAs these are the bandgap energies $E_\mathrm{g}$, electron mass $m_\mathrm{e}/m_0$, and hole masses $m_\mathrm{hh}/m_0$ and $m_\mathrm{lh}/m_0$ calculated from the $\gamma$-values. 
The bandgap energies and electron/hole masses for the alloys (In,Ga)As and (In,Al)As are linearly interpolated, using respective bowing parameters~\cite{Vurgaftman2001}. 
The valence band and conduction band offsets ($\mathrm{VBO} = \SI{26.8}{\percent}$, $\mathrm{CBO} = \SI{73.2}{\percent}$) are based on the recommended values for the well-studied composition $\mathrm{In}_{0.53}\mathrm{Ga}_{0.47}\mathrm{As/In}_{0.52}\mathrm{Al}_{0.48}\mathrm{As}$~\cite{Vurgaftman2001}.

\begin{table}[]
	\caption{Parameters of constituents used in the (In,Ga)As/(In,Al)As QW calculations (recommended values from Ref.~\cite{Vurgaftman2001}) and resulting parameters for the QW materials.}
	\label{tab:materials-parameters}
	\begin{ruledtabular}
	\begin{tabular}{ccccccc}
		 	& $E_\mathrm{g}$ (eV) & $\gamma_1$ & $\gamma_2$ & $m_\mathrm{hh}/m_0$ & $m_\mathrm{lh}/m_0$ & $m_\mathrm{e}/m_0$ \\ \hline
	InAs 	& $0.417$             & $20.0$       & $8.5$        & $0.333$         & $0.027$         & $0.026$		\\
	GaAs 	& $1.519$             & $6.98$       & $2.06$       & $0.350$         & $0.090$         & $0.067$		\\
	AlAs 	& $3.099$             & $3.76$       & $0.82$       & $0.472$         & $0.185$         & $0.150$		\\ \hline
	InGaAs	& $0.733$			  &	-			 & -			& $0.374$		  & $0.047$		    & $0.040$		\\	
	InAlAs	& $1.322$			  &	-			 & -			& $0.389$		  & $0.090$		 	& $0.064$
	\end{tabular}
	\end{ruledtabular}
\end{table}

Substituting the above values of $\Delta_\mathrm{l,V} = \SI{0.15}{meV}$ and $\Delta_{\rm lh}= 19$~meV into the Eq.~\eqref{eq:Pc} we obtain $P_\mathrm{c} \approx \SI{0.82}{\percent}$ which can be used for the final calculation of the directionality in the hybrid plasmonic-semiconductor QW structure. 
The theoretical model based on the scattering matrix method presented above (see Fig.~\ref{fig:5}(c)) and in Ref.~\onlinecite{FelixNature-2018} allows us to predict the spectral and angular dependence of the magnetic field induced changes of the emission $\rho(E,\theta)$.
This is done by first calculating the emission for oppositely polarized emitters in the quantum well (i.e.\ opposite external magnetic fields) and then $\rho$ according to Eq.~\eqref{eq:rho}, which is equivalent to $C$.
We use a hybrid structure with a plasmonic grating similar to that in our experimental studies with grating period $a = \SI{420}{nm}$, slit width $w = \SI{70}{nm}$ and gold stripe thickness of $\SI{40}{nm}$, located $\SI{30}{nm}$ from the (In,Ga)As QW by an (In,Al)As cap layer. 
The buffer layer on the other side of the QW is $\SI{5}{\micro\m}$ thick and GaAs is used as substrate.
Like for the previously presented simulations the energy-dependent refractive indices for gold and GaAs were taken from Refs.~\onlinecite{Aspnes1986-n_GaAs, Johnson1972-n_Au}, whereas the refractive index for the (In,Al)As buffer was assumed as $n=\num{3.3}$ based on a similar composition examined in Refs.~\onlinecite{Dinges1992-n_InAlAs, Linnik2002-n_InAlAs}.

Figure~\ref{fig:Simu} summarizes the modeled angular and spectral dependences of the emitted signal and its directionality in two-dimensional plots. 
The mean intensity emitted via the plasmonic gold grating is calculated as $I_0=(I(+B)+I(-B))/2$ and $C = \rho$ is obtained using Eq.~\eqref{eq:rho}.
It is assumed that optical transitions with $P_\mathrm{c} = 0.82\%$ contribute equally and independently in the full spectral range of interest ($0.7-1.0$~eV). 
Therefore, the intensity plot shown in Fig.~\ref{fig:Simu}(a) should be considered rather as a response function of the hybrid structure, which should be further multiplied with the corresponding PL spectrum of the particular QW structure. 
This assumption is valid as long as we can neglect the contribution from light holes, i.e.\ $\Delta_{\rm lh} \geq  k_\mathrm{B}T_\mathrm{c}$. 
The lines with high intensity in Fig.~\ref{fig:Simu}(a) follow the dispersion branches of the SPPs at the interface between metal and semiconductor. 
If we consider SPPs with positive wavevector, i.e.\ positive emission angle $\theta$, the lower (higher) energy branch in the spectral range of $0.7$ to $0.85$~eV ($0.9$ to $1.0$~eV) corresponds to SPPs with negative (positive) group velocity.
For negative $\theta$ the situation is inverse. 
Thus, for a given emission angle, SPP propagation and consequently its circular polarization in the $yz$-plane are opposite for the high and low energy modes (see also Fig.~\ref{fig:1}(b)).
The angular and spectral dependences of directionality shown in Fig.~\ref{fig:Simu}(b) correlate with the position of the SPP resonances in Fig.~\ref{fig:Simu}(a). 
For a given angle $\theta$ there is a sign change in $C$ when the photon energy of emission is scanned from the low to the high energy SPP mode. 
As expected, the change of the sign is attributed to coupling of the emission to SPPs propagating in opposite directions. 
The directionality reaches its maximum in the vicinity of the SPP resonances with $C \approx \SI{5}{\percent}$. 
Note that there is a further possibility to tune the directional emission for specific emission angles and energies by modifying the gold grating parameters and thus moving the plasmonic resonance through the spectrum. 

\begin{figure}
	\includegraphics[width=0.48\textwidth]{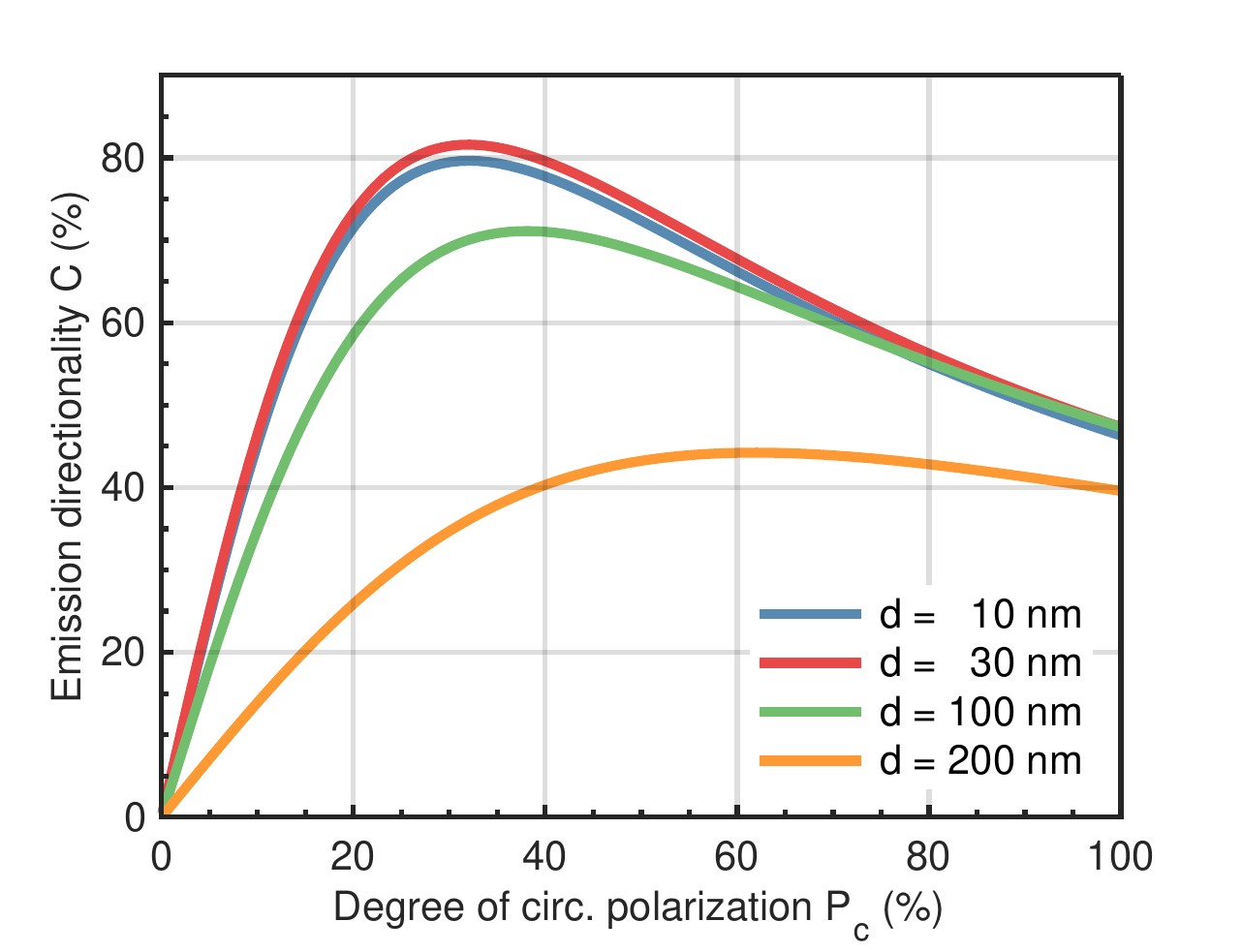}
	\caption{
		Dependence of the directionality $C$ on the degree of circular polarization $P_\mathrm{c}$ for different cap layer thicknesses $d$ in the hybrid (In,Ga)As/(In,Al)As QW structure at photon energy $E = \SI{0.8}{eV}$ and emission angle $\theta=\SI{15}{\degree}$. 
		There is a maximum at $d = \SI{30}{nm}$, likely corresponding to the emitter polarization matching the SPP polarization.
		For larger distances the interaction between the QW emission and the evanescent SPP modes decreases.}
	\label{fig:C-from-P_c}
\end{figure}

The results of Fig.~\ref{fig:Simu} demonstrate that routing up to $C = \SI{5}{\percent}$ for $P_\mathrm{c} = \SI{0.82}{\percent}$ in the temperature range up to $\SI{200}{K}$ is possible, enabling the application at temperatures reachable with cooling via the Peltier effect. 
In general by decreasing the energy splitting $\Delta_{\rm lh}$ the polarization $P_\mathrm{c}$ can be increased, but the operation range will be limited to lower temperatures. 
Similarly, a larger splitting $\Delta_{\rm lh}$ extends the effect to larger temperatures but decreases the achievable $P_\mathrm{c}$ and consequently $C$. 
In Fig.~\ref{fig:C-from-P_c} we present the results of calculations for the dependence of the emission directionality for $\theta=15^\circ$ and $E = \SI{0.8}{eV}$ on the degree of circular polarization in the QW $P_\mathrm{c}$. 
The results are presented for different distances $d$ between the circular dipole emitter and the plasmonic interface. 
For $P_\mathrm{c} < \SI{20}{\percent}$ the emission directionality $C$ increases linearly with $P_\mathrm{c}$. 
For higher $P_\mathrm{c}$ the directionality $C$ reaches a maximum and slowly decreases then. 
Regarding the distance between QW and surface $d$, an optimum exists at $d = \SI{30}{nm}$, which is most probably corresponding to the best matching between the emitter polarization and polarization of the propagating SPP. 
For larger distances the interaction between the QW emission and the interface bound SPPs decreases, which is due to the evanescent nature of SPPs, as it was also shown experimentally in Ref.~\onlinecite{FelixNature-2018}. 
We note that compared to (Cd,Mn)Te based structures, in the (In,Ga)As QW there is a significant increase in the wavelength of light emission. 
Therefore, the corresponding distances for $d$ where SPP enhancement is present increases as compared to the previous case. 
In addition the propagation length of SPPs at the flat interface between semiconductor and metal increases by an order of magnitude which can be used for applications where the routing should be accomplished directly on chip. 

We also note that another well-studied lattice matched semiconductor heterostructure worth investigating could be an $\mathrm{In}_{0.53}\mathrm{Ga}_{0.47}\mathrm{As}$ QW with InP barriers~\cite{Vurgaftman2001}. 
The main difference compared to the previously discussed system are the different offsets of the valence and conduction bands of this heterojunction with a larger valence band offset for (In,Ga)As/InP, creating a stronger confinement of the holes in the QW and thus possibly a larger hole g-factor.

\section{Conclusions}
We have studied the temperature dependence of transverse magnetic routing of light emission in hybrid plasmonic-semiconductor QW structures. Coupling of the QW emission into the surface plasmon polariton modes, which are supported by a one-dimensional gold grating in direct vicinity of the QW, allows one to enhance the directional emission significantly. One of the key features which determines the magnitude of TMRLE is the degree of circular polarization for interband optical transitions in the plane perpendicular to the magnetic field. The degree of circular polarization is determined by the competition of heavy-light hole mixing by magnetic field and the energy splitting $\Delta_\mathrm{lh}$ between the upper heavy- and light-hole subbands due to confinement. It grows linearly with magnetic field strength, as it is proportional to the Zeeman energy of light holes in transverse magnetic field $\Delta_\mathrm{l,V}$, and inversely proportional to $\Delta_\mathrm{lh}$. Essentially the magnitude of circular polarization for optical transitions associated with heavy- and light-holes is the same while their helicities are opposite. This consequently leads to opposite directionality contributions which compensate each other. As a result the population of the light hole subband at higher temperatures plays an important role in the total directionality of the emission. The routing is diminished as soon as the thermal energy exceeds the splitting between the heavy and light holes.

We have compared the magnitude and temperature dependence of TMRLE in hybrid nanostructures based on different semiconductor constituents. In the diluted magnetic semiconductor (Cd,Mn)Te/(Cd,Mg)Te QW structure a strong directionality up to \SI{15}{\percent} at low temperatures of about \SI{20}{K} and magnetic field strength of \SI{0.5}{T} is measured. We observe that the degree of directionality in (Cd,Mn)Te based QW structures decreases with increasing temperature due to reduction of the Zeeman splitting $\Delta_\mathrm{l,V}$, which is proportional to the manganese polarization and follows the modified Brillouin function, i.e.\ the magnetic susceptibility of the Mn system. The contribution from optical transitions associated with the light holes with opposite sign has been demonstrated experimentally. To make the routing magnitude robust against temperature changes we propose nonmagnetic narrow band gap semiconductor QW structures with intrinsically large g-factor. In particular we modeled the TMRLE in (In,Ga)As/(In,Al)As heterostructures where the nearly temperature independent Zeeman splitting results in noticeable routing with a directionality degree of \SI{5}{\percent} in the temperature range up to \SI{200}{K} for a magnetic field of \SI{1}{T}.

The design and variation of parameters of the plasmonic constituent opens further possibilities for investigation and optimization of TMRLE. In case of one-dimensional gratings the period and slit width can be used to tailor the SPP dispersion in order to achieve fine tuning of the angular distribution of the directional emission as well as its magnitude. Moreover modifications of the energy spectrum in the QW due to coupling between excitons and plasmons~\cite{Vasa2008} and its impact on TMRLE require further investigation. 

In contrast to the typical situation where the heavy-hole state is the ground state, in quantum well structures like ZnSe/(Cd,Zn)Se the light-hole state has the lowest energy. 
Qualitatively both cases should show the same results (see Fig.~\ref{fig:pola2}).
Nevertheless it is worth investigating, also because the light holes can be critical for other phenomena like optical orientation~\cite{Akimov-Vondran-PRB}, which is the opposite process to the TMRLE presented here.

\begin{acknowledgments}
We are grateful to M.~V.~Durnev, A.~A.~Toropov and V.~L.~Korenev for useful discussions.
We acknowledge the financial support by the Deutsche Forschungsgemeinschaft through the International Collaborative Research Centre 160 (Project C5). 
ANP acknowledges partial financial support from the Russian Foundation for Basic Research Grant No.19-52-12038-NNIO a. 
The research in Poland was partially supported by the Foundation for Polish Science through the IRA Programme co-financed by EU within SG OP (Grant No. MAB/2017/1) and by the National Science Centre through Grant No. 2018/30/M/ST3/00276.
\end{acknowledgments}

\appendix*	
\section{Magneto-PL in Faraday geometry}
\label{sec:Appendix_MagnetoPL}

\begin{figure}
	\includegraphics[width=0.48\textwidth]{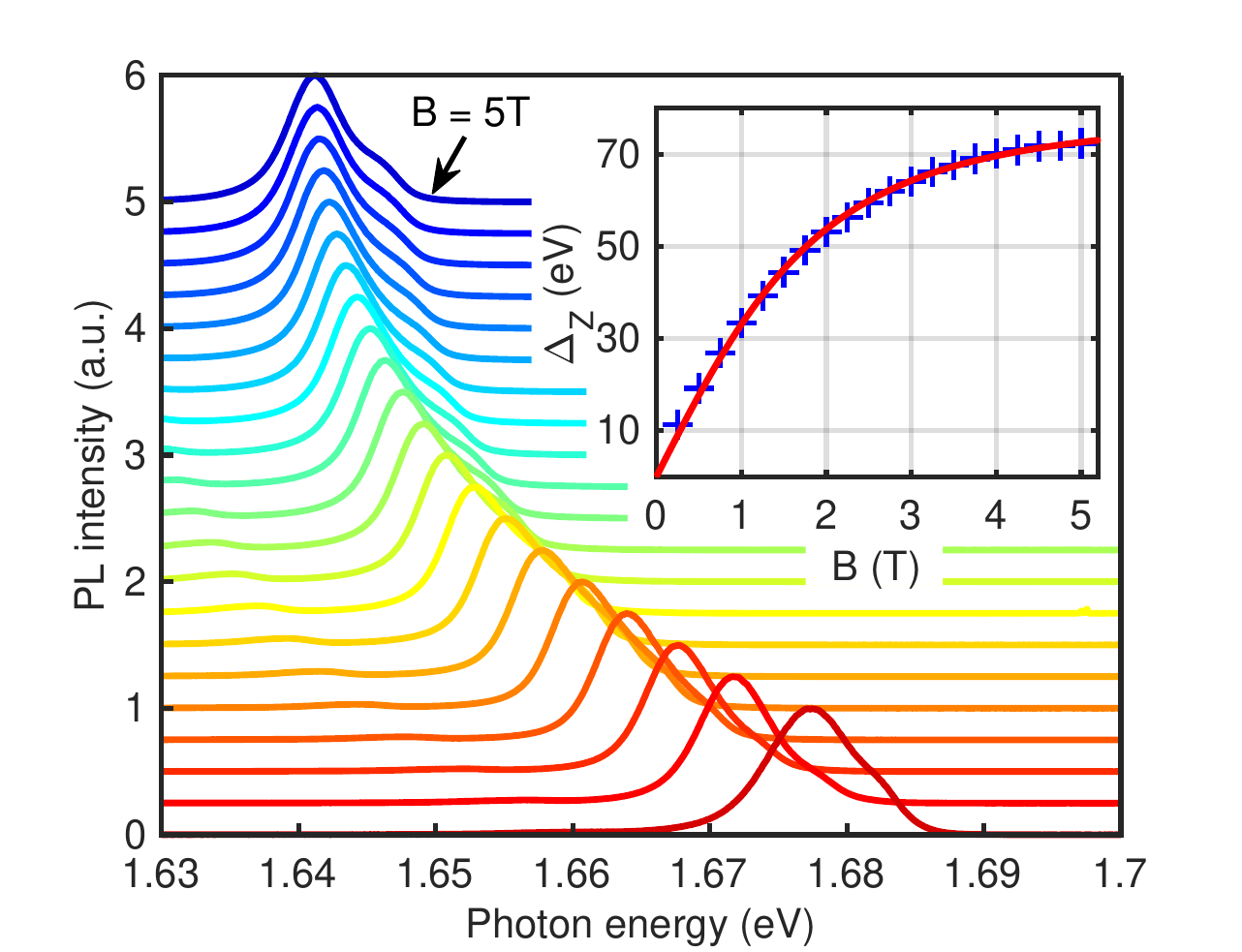}
	\caption{
			Spectra of PL from the DMS (Cd,Mn)Te/(Cd,Mg)Te QW in arbitrary units for magnetic fields in Faraday geometry between \SI{0}{T} (bottom, red) and \SI{5}{T} (top, blue) in \SI{250}{mT} steps.
			The energy of the PL maximum shifts to lower energies with increasing field due to the Zeeman splitting. 
			The interception with the y-axis also depicts the B-field for each graph.
			Inset: Zeeman splitting of the heavy hole exciton $\Delta_{\mathrm{Z}}$ (Eq.~\eqref{eq:Zeeman_experiment}) for different external magnetic fields $B$ in Faraday geometry and Eq.~\eqref{eq:Zeeman_Faraday} fitted to these data.}
	\label{fig:AppendixZeeman}
\end{figure}

The Zeeman splitting of the heavy-hole exciton energy levels in Faraday geometry $\Delta_{\mathrm{Z}}$ allows one to retrieve structure parameters like the manganese concentration $x$ and effective temperature $T_{0}$ from magnetic field dependent PL measurements.
For this the (Cd,Mn)Te/(Cd,Mg)Te QW sample (see Sec.~\ref{sec:Experimental}) was mounted in a pumped liquid helium bath cryostat at $T = \SI{1.5}{K}$ with superconducting coils allowing magnetic fields up to $B=\SI{\pm5}{T}$ in Faraday geometry.
Linearly polarized light from a Ti:Sa-laser with $\lambda = \SI{690}{nm}$ ($E = \SI{1.797}{eV}$) is focused onto the sample without plasmonic grating at a small angle using a $f = \SI{200}{mm}$ lens.
A $f = \SI{400}{mm}$ lens collects and collimates the light emitted from the sample in reflection geometry.
The combination of a quarter-wave plate and a Glan-Thompson prism are used to select either $\sigma^{+}$ or $\sigma^{-}$ polarized light, which is then focused onto the slit of a spectrometer by another lens.
There the light is split into its spectral components and measured by a nitrogen-cooled CCD camera. 
To obtain the Zeeman splitting of the PL from the QW we measured the spectrum of the emitted light for magnetic fields between $\SI{\pm5}{T}$ in steps of $\SI{0.25}{T}$ separately for $\sigma^{+}$ and $\sigma^{-}$ polarization at each step. 
Figure~\ref{fig:AppendixZeeman} shows the spectra of the PL from the QW in arbitrary units for different magnetic fields, highlighting the shift of the PL maximum due to the applied magnetic field.
For $\sigma^{+}$ polarization the emission maximum is located at $\SI{1.677}{eV}$ for $B=0$, shifting to lower energies and down to $\SI{1.641}{eV}$ at $\SI{5}{T}$.

Due to the giant Zeeman splitting in DMS materials the emission is strongly polarized and already for moderate magnetic fields only one circular polarization is detectable.
Therefore, the Zeeman splitting in DMS can be calculated as twice the energy difference between $B = 0$ and $B>0$
\begin{equation}
	\Delta_{\mathrm{Z}} = 2\left(E_{\sigma^{+}}(B=0) - E_{\sigma^{+}}(B>0)\right)\text{,}
	\label{eq:Zeeman_experiment}
\end{equation}
which yields $\SI{72.4}{meV}$ at $\SI{5}{T}$.
The Zeeman splitting for external magnetic fields between \SI{0}{T} and \SI{5}{T} is shown as inset in Fig.~\ref{fig:AppendixZeeman}.
It increases with magnetic field and reaches almost saturation value at $B = \SI{5}{T}$.

Similar to Eq.~\eqref{eq:Zeeman} the heavy-hole exciton energy level splitting of the DMS sample can be written as \cite{Ossau-1993}
\begin{equation}
	\Delta_{\mathrm{Z}} = xN_{0}(\alpha-\beta)\left<S_{z}^\mathrm{Mn}\right> 
	\label{eq:Zeeman_Faraday}
\end{equation}
with the Mn concentration $x$, the exchange constants for the conduction band $N_{0}\alpha = \SI{0.22}{eV}$ and valence band $N_{0}\beta = \SI{-0.88}{eV}$, and $\left< S_z^\mathrm{Mn}(B)\right>$ the thermal average of the ${\rm Mn^{2+}}$ spin projection along the external magnetic field according to Eq.~\eqref{eq:Brillouin}~\cite{Ossau-1993}.
Using Eq.~\eqref{eq:Zeeman_Faraday} to fit the experimental data yields the curve shown in the inset of Fig.~\ref{fig:AppendixZeeman} and the parameters $T_0 = \SI{1.9\pm0.1}{K}$ and $xS_{\mathrm{eff}} = \num{0.071\pm0.001}$.
$xS_{\mathrm{eff}}$ is used as fitting parameter as the exact Mn content of our sample is unknown.
However, it can be estimated to $x \approx \SI{4}{\percent}$ using the experimental data from Ref.~\cite{Ossau-1993}, which corresponds to $S_\mathrm{eff} \approx \num{1.7}$.

The exciton Zeeman splitting in Faraday geometry $\Delta_\mathrm{Z}$ measured here is connected to the Zeeman splitting of the light holes in Voigt geometry $\Delta_\mathrm{l,V}$ used for the description of TMRLE via $P_\mathrm{c}$ in Eq.~\eqref{eq:Pc} via
\begin{equation}
	\Delta_\mathrm{l,V} = \frac{2}{3}\Delta_\mathrm{h,F} = \frac{2}{3}\frac{\left|\beta\right|}{\left|\alpha-\beta\right|}\Delta_\mathrm{Z}\text{.}
\end{equation}


\end{document}